\documentclass[useAMS,usegraphicx,usenatbib,letterpaper]{mn2e}
\usepackage{epsfig}
\usepackage{longtable}
\usepackage{graphics}
\usepackage{times}
\usepackage{epstopdf}
\usepackage[dvips]{graphicx}
\usepackage{latexsym}
\usepackage{amsmath}
\usepackage{amsfonts}
\usepackage{amssymb}
\usepackage{multicol}
\usepackage{array}
\usepackage{multirow}
\usepackage{subfigure}
\usepackage{caption}

\def\lea{\mathrel{<\kern-1.0em\lower0.9ex\hbox{$\sim$}}}
\def\gea{\mathrel{>\kern-1.0em\lower0.9ex\hbox{$\sim$}}}
\def\leq{\mathrel{<\kern-1.0em\lower0.9ex\hbox{$-$}}}
\def\geq{\mathrel{>\kern-1.0em\lower0.9ex\hbox{$-$}}}

\title[Coma Dwarf Galaxies: I. Velocity Dispersion
Measurements]{Dwarf Galaxies in the Coma Cluster: I. Velocity Dispersion
Measurements\thanks{Based in part on observations made with the NASA/ESA Hubble Space Telescope, obtained at the Space Telescope Science Institute, which is operated by the Association of Universities for Research in Astronomy, Inc. under  NASA contract  NAS 5-26555. These observations are associated with programme GO10861.}\thanks{Some of the data presented herein were obtained at the W.M. Keck Observatory, which is operated as a scientific partnership among the California Institute of Technology, the University of California and the National Aeronautics and Space Administration. The Observatory was made possible by the generous financial support of the W.M. Keck Foundation. }}

\author[Kourkchi et al.]  {E. Kourkchi$^{1,2}$, H. G. Khosroshahi$^{1}$,
D. Carter$^{3}$, A.M. Karick$^{3}$, E. M\'armol-Queralt\'o$^{4}$, \newauthor
K. Chiboucas$^{5,6}$, R. B. Tully$^{5}$, B. Mobasher$^{7}$, R. Guzm\'an$^{8}$, A. Matkovi\'{c}$^{9}$ and N. Gruel$^{10}$ \\$^1$ School of Astronomy, Institute for Research in Fundamental Sciences (IPM), PO Box 19395-5531,Tehran, Iran.  \\$^2$ Department of Physics, Sharif University of Technology, P.O.Box:11155-9161, Tehran, Iran.  \\$^3$ Astrophysics Research Institute, Liverpool John Moores University, Twelve Quays House, Egerton Wharf, Birkenhead CH41 1LD, UK.  \\$^4$Departamento de Astrof\'sica, Facultad de Ciencias F\'isicas, Universidad Complutense de Madrid, 28040 Madrid, Spain. \\$^5$Institute for Astronomy, University of Hawaii, 2680 Woodlawn Dr., Honolulu, HI 96821, USA. \\$^6$Gemini Observatory, Northern Operations Center, 670 N. A'ohoku Place, Hilo, Hawaii, 96720, USA. \\$^7$Department of Physics and Astronomy, University of California, Riverside, CA 92521, USA. \\$^8$Department of Astronomy, University of Florida, PO. Box 112055, Gainsville, FL 32611-2055, USA. \\$^9$Department of Astronomy and Astrophysics, Pennsylvania State University, University Park, PA 16802, U.S.A. \\$^{10}$Centro de Estudios de F\'{i}sica del Cosmos de Arag\'{o}n, C. General Pizarro, 1-3 44001 Teruel, Spain.
 }

\begin{document}

\date{Accepted 2011 September 27.  Received 2011 September 13; in original form 2010 December 18}
\pagerange{\pageref{firstpage}--\pageref{lastpage}}
\pubyear{2010} \volume{000} \pagerange{1}

\maketitle 

\label{firstpage}

\begin{abstract}
We present the study of a large sample of early-type dwarf galaxies in the Coma cluster observed with DEIMOS on the Keck II to determine their internal velocity dispersion. We focus on a subsample of 41 member dwarf elliptical galaxies for which the velocity dispersion can be reliably measured, 26 of which were studied for the first time. The magnitude range of our sample is $-21<M_R<-15$ mag.

This paper (paper I) focuses on the measurement of the velocity dispersion and their error estimates. The measurements were performed using {\it pPXF (penalised PiXel Fitting)} and using the Calcium triplet absorption lines. We use Monte Carlo bootstrapping to study various sources of uncertainty in our measurements, namely statistical uncertainty, template mismatch and other systematics. We find that the main source of uncertainty is the template mismatch effect which is reduced by using templates with a range of spectral types.

Combining our measurements with those from the literature, we study the Faber-Jackson relation ($L\propto\sigma^\alpha$) and find that the slope of the relation is $\alpha=1.99\pm0.14$ for galaxies brighter than $M_R\simeq-16$ mag. A comprehensive analysis of the results combined with the photometric properties of these galaxies is reported in paper II.

\end{abstract}

\begin{keywords}
galaxies: clusters: individual: Coma; galaxies: elliptical and lenticular, cD; galaxies: dwarf; galaxies: kinematics and dynamics; galaxies: fundamental parameters; galaxies: evolution
\end{keywords}

%-----------------------------------------------------------------------------
\section{Introduction}
\label{chap:introduction}

A clear understanding of dwarf galaxies and their relation to their most massive counterparts is essential for testing galaxy formation models. Although the lowest luminosity dwarf galaxies ($M_v \sim -9$) appear to contain large amounts of dark matter (Aaronson et al. 1983; Mateo 1998) for brighter dwarfs the situation is less clear. De Rijcke et al. (2006)  find that the dwarf companions to M31, at $M_v \sim -14$, contain around 40-50\% dark matter by mass within the inner two half-light radii. Geha et al (2002) find that six Virgo cluster dwarfs at $M_v \sim -16$ do not show evidence for dark matter within an effective radius. Toloba et al. (2011) observe a larger sample of Virgo dwarfs, and again find that they are not dark matter dominated. Here we observe a sample of Coma cluster dwarfs in a somewhat brighter magnitude range ($−21 < M_R < −15$) in order to understand better their internal dynamics. Our goals are the extension of the Faber-Jackson relation ($L \propto \sigma^{\alpha}$) and the fundamental plane (FP, Djorgovski \& Davis 1987; Dressler et al. 1987; Bender et al. 1992) relating luminosity, velocity dispersion, surface brightness, and scale length of galaxies, to the low luminosity dwarf population, providing important tools for studying formation of galaxies. For example, the precise value of the exponent of the Faber-Jackson relation, $\alpha$, helps to constrain galaxy formation models as it is sensitive to the effect of gas loss from self-gravitating systems (Dekel \& Silk 1986;  Yoshii \& Arimoto 1987). In the case where a dark matter halo governs the galaxy dynamics, $\alpha$ is predicted to be 5.26 and $M/L$~$\propto$~$L^{−0.37}$. If, instead, the galaxies contain only baryonic matter and have roughly constant $M/L$, a flatter relation, $L\propto\sigma^{2.7}$, is expected (Dekel \& Silk 1986).

Clusters of galaxies are ideal for studying scaling relations as they provide large number of galaxies of different types, at a common distance, different environments within the cluster, and allow one  to assess the interaction with the Intra-Cluster Medium (ICM). Since the Coma cluster is one of the nearest, rich and dense clusters, it has been a popular target for studies of scaling relations among galaxies including the  Faber-Jackson relation (J{\o}rgensen et al. 1996; Moore et al. (2002: MLKC02); Graham \& Guzm\'an 2003).
 
The Faber-Jackson relation for faint ellipticals in Coma cluster has been studied by Matkovi\'c \& Guzm\'an (2005: MG05) and Cody et al. (2009: Co09), using WIYN/HYDRA (multi-fiber) spectroscopy. Both MG05 and Co09 find $L_R \propto \sigma^{2.0\pm0.3}$ for dwarf ellipticals even though these studies have only 10 galaxies in common whose $\sigma$ measurements differ by up to $\sim$30\% for the faintest galaxies. Thus, it is important to obtain high S/N data to measure velocity dispersions for dwarf galaxies at fainter magnitudes.

In this paper we use the high spectral resolution of DEIMOS (Faber et al. 2003) on the Keck II telescope to measure the velocity dispersions for a sample of 41 faint elliptical galaxies in the core of the Coma cluster. Of these galaxies, 15 are common with the samples of MG05, Co09 or both, and for the remaining 26, these are the first velocity dispersion measurements. The current DEIMOS measurements extend the study to luminosities in the range $-16.5 < M_R < -15$, about 1 magnitude fainter than the limit reached by the WIYN/HYDRA data in previous studies of Coma dwarf galaxies.

This paper covers mainly the observations and the technical part of the analysis. The emphasis is on the reliable estimate of the uncertainties in measuring the internal velocity dispersion of galaxies.  In section 2 we describe the observations, the spectroscopic setup and the data reduction. Section 3 is dedicated to the measurement of radial velocities and the velocity dispersions. The error analysis is described in section 4. The results of the analysis and the Faber-Jackson relation are presented and discussed in section 5. 

Throughout this paper, the distance modulus of the Coma cluster is considered to be $35.00~mag$ (Carter et al. 2008).

\label{introduction}

%-----------------------------------------------------------------------------
\section{Observations}
\subsection{Source Selection}
\label{selection}

Galaxies were selected using the photometry from Adami et al. (2006). The (B-R) versus R colour-magnitude relation was constructed, and the red sequence identified and fitted by a linear relation. Galaxies were then selected from within a region $\pm 3\sigma$ from this relation, with a magnitude range $-19 < M_R < -16, ~or~ 16 < R < 19$. Images from the  HST/ACS Coma Treasury Survey (Carter et al. 2008) were then used to reject a very small number of galaxies with obvious spiral structure. However not all candidates fell within the footprint of the completed survey (which was truncated owing to ACS failure) so some late-type galaxies may remain in the sample. The masks were then designed at the university of Hawaii to accommodate, and as a compromise between the two programmes: the velocity dispersions programme described here, and a separate programme to measure redshifts of a sample of low surface brightness galaxies. The requirement that the spectra from different slits should not overlap, and the need for sufficient slit length to enable sky subtraction, places considerable restriction on the number of galaxies one can observe in each mask.

%-----------------------------------------------------------------------------
%-----------------------------------------------------------------------------

\begin{figure*}
\begin{center}
\includegraphics[width=15cm]{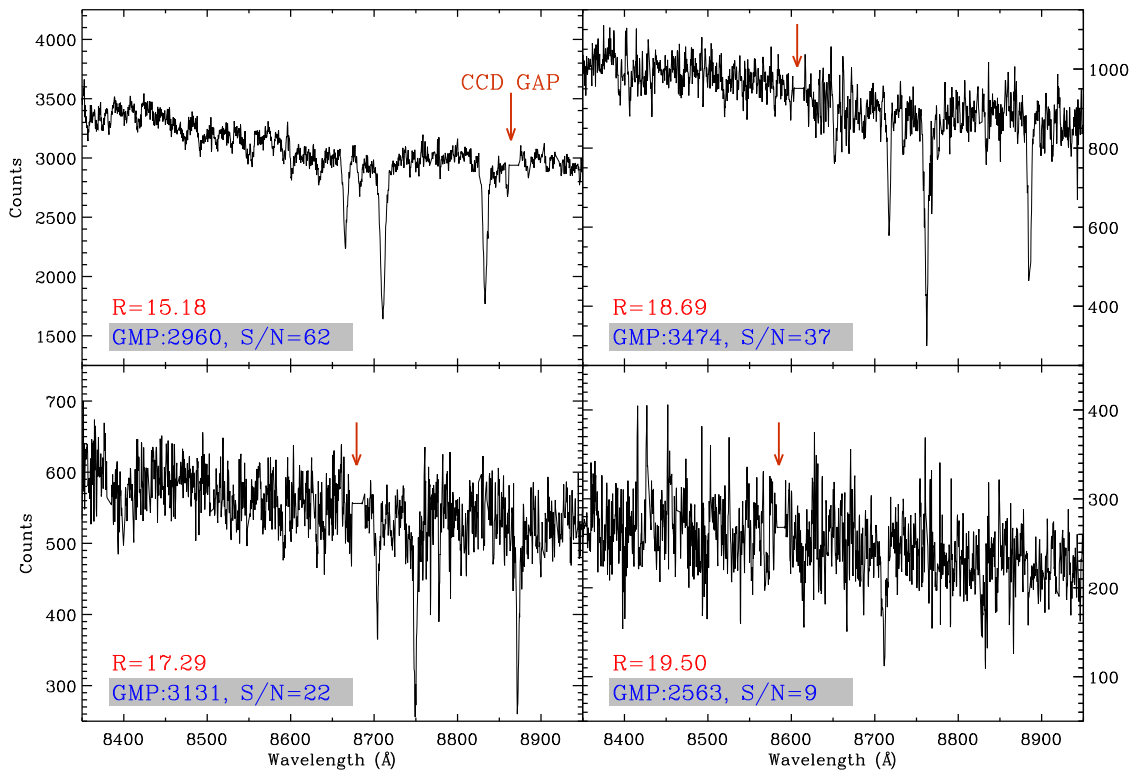}
\end{center}
\caption{
Example of galaxy spectra in the sample with different values of signal-to-noise ratio (S/N). All S/N values are estimated per pixel. The Godwin, Metcalfe $\&$ Peach (1983; GMP) ID and R magnitude are given for each galaxy. In each panel, the arrow shows the location of CCD gap in the spectrum.} 
\label{fig:spectra}
\end{figure*}

\subsection{Spectroscopic setup}

\label{chap:spectra}

The observations were carried out on the night of March 19$^{th}$ 2007, using DEIMOS on the Keck II telescope. The outer dimension of the rectangular DEIMOS field is 16.7$\times$5.0 arcmin$^2$ and all observed spectra are mapped onto 2$\times$4 mosaic of 2K$\times$4K CCDs (2 CCDs for each spectrum). We used 1200G grating (1200 grooves/mm BK7, gold-coated, and blazed at 7760 \AA). The nominal central wavelength at the centre of the mask was 8700 \AA, although of course the central wavelength of a given spectrum depends upon the position of that slitlet in the mask. Although slitlets for the redshift survey observations were drilled at a variety of widths, all of those for the velocity dispersion project described here had widths of 0.7 arcsec. Reciprocal dispersion is 0.33 \AA/pixel, giving a spectral resolution (FWHM) of 1.6 \AA, and a resolving power R$\sim$5000.  This equates to an instrumental velocity width of $\sim$25 km s$^{-1}$, allowing us to measure velocity dispersions down to $\sim$15 km s$^{-1}$ in spectra with S/N $\ge 15$ per pixel. The velocity dispersions of galaxies in the $17 \le R \le 19$ dE sample are expected to range between 10 km s$^{-1}$ and 80 km s$^{-1}$. The spectral range covered is approximately $7500-10,000~\AA$, although it depends upon the slit position in the mask. This wavelength range is dominated by the very deep calcium II triplet absorption lines (CaT) at $8498.02, 8542.09$ and $8662.14~ \AA$, which depend only weakly upon metallicity, and therefore provide an ideal range over which to measure the kinematic properties of low-luminosity galaxies. A GG495 filter was used to block out the contamination from the second order blue spectrum.

Two masks were observed for each of two pointings, and the exposure time for each mask was in the range 6030 to 6600 seconds. A summary of these observations is presented in Table ~\ref{tab:observations}. In addition, a number of potential template stars, and flux calibration stars, were observed using a 0.7 arcsec width long slit during twilight at the beginning and end of the night. Wavelength calibration arcs and observations of internal flatfielding lamps were also made at the beginning of the night.

We observed eleven standard stellar spectra of which only two templates were reliable for measurements in this study. The other stellar templates were excluded because they either were not G/K-type stars or were affected by noise near the Ca triplet lines.

\subsection{Data reduction}
\label{chap:reduction}

The spectra of dwarf galaxies and template stars were reduced using the IDL {\it spec2d } pipeline developed by the DEEP2 Galaxy Redshift Survey team at the University of California-Berkeley for that project (see Davis et al. (2003) for more information). The pipeline performs the flat-fielding using internal quartz flats and the wavelength calibration using the ArKrNeXe arc lamps. The sky subtraction and  the data extraction to 1D spectra are also done within the pipeline. For our purpose here, all spectra are not flux calibrated and all partially remaining sky emission lines, particularly near the Ca triplet lines, were manually removed. In this manual process, for each galaxy spectrum, we checked the wavelength of the remaining bright emission lines with the catalogues of the sky emission lines and then we replaced the verified sky emission lines with the continuum fitted underneath the spectrum. Each spectrum was originally located on 2 CCDs. In order  to join the blue and red parts of each 1D spectrum, the average value of the nearest 50 pixels to the CCDs gap was calculated and the offset was used to construct the conjoined thorough 1D spectrum. An example of final one-dimensional spectra from the DEIMOS in the range $\sim$8300-9000 \AA~ is shown in Figure \ref{fig:spectra}. In this paper all spectral S/N values are estimated per pixel (i.e. 0.33 \AA).

\begin{figure*}
\centering
\includegraphics[width=12cm]{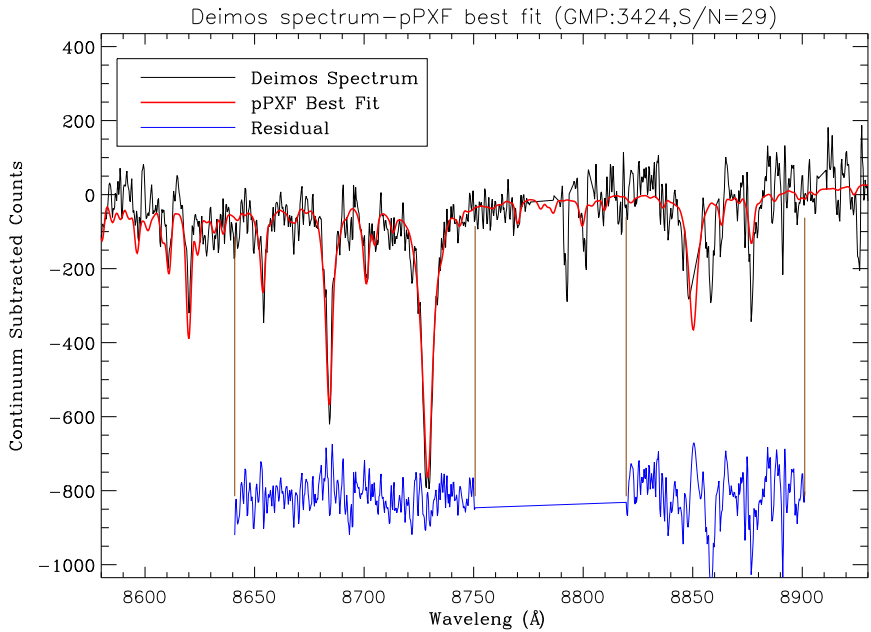}
\caption {An Example of pPXF fit to a galaxy (GMP3424) from the 2007 DEIMOS run. Thin black is the observed spectrum, and the thick black (red) is the best pPXF fit using a single stellar template. The residuals over the fitted region is also shown (blue). The S/N for this spectrum is 29, and we find $V_r = 5596\pm 7 km~s^{-1}$; $\sigma = 27\pm 9 km~s^{-1}$.}
\label{fig:ppxf}
\end{figure*}

\begin{table}
\begin{center}
\renewcommand{\arraystretch}{0.99}\renewcommand{\tabcolsep}{0.12cm}
\caption{List of the exposures on the galaxy masks. In column 5, we give the number of slits drilled at 0.7 arcsec width, specifically for the velocity dispersion programme, excluding those drilled at larger widths for redshift measurements of low surface brightness galaxies. In column 6, we give the total number of slits per each mask. }
\begin{tabular}{c c c c c c}\hline
Mask&\multicolumn{2}{c}{Mask Centre (J2000)}&Exposure&\multicolumn{2}{c}{No. of Slits}  \\ 
&RA &Dec&Time (s)& dEs & Total \\ 
(1) & (2) & (3) & (4) &  (5) & (6) \\
\hline
Coma1-1&12:59:57.83&27:59:30.6&6600&24 & 105\\
Coma1-2&12:59:57.83&27:59:31.2&6600&15 & 109 \\
Coma2-1&13:00:04.04&27:54:52.2&6600&22  & 101 \\
Coma2-2&13:00:04.04&27:54:52.2&6030&14  & 103 \\
\hline
\label{tab:observations}
\end{tabular}
\end{center}
\end{table}

%\section{Spectroscopic Measurement}
%\label{chap:specanalysis}
%-----------------------------------------------------------------------------
\section{The Measurement of Radial Velocity and Velocity Dispersion} 
\label{chap:velocities}

There are 49 galaxies in our DEIMOS sample with spectroscopically confirmed Coma membership (Marzke et al. 2011). The CaT absorption lines for eight galaxies were either coincident with gaps in the CCD mosaic, or too low signal-to-noise to make reliable measurements and therefore were taken out of the sample. We thus, used 41 spectra with identified CaT lines with signal-to-noise ratio (S/N) ranging from 5 to 87. Figure~\ref{fig:spectra} shows four of the sample spectra.

The line-of-sight radial velocities, and velocity dispersions, $\sigma$, were measured using the pPXF\footnote{penalized PiXel Fitting} software developed by Cappellari $\&$ Emsellem (2004). The pPXF technique works in pixel space and uses Gauss-Hermite series to extract the radial velocity and velocity dispersion simultaneously by minimizing the $\chi^2$ which is defined as

\begin{equation}
\chi^2=\sum_{n=1}^N \frac{G_{mod}(x_n)-G(x_n)}{\Delta G(x_n)}, 
\end{equation}

where $G(x_n)$ and $G_{mod}(x_n)$ are the original and the modelled galaxy spectra respectively. $\Delta G(x_n)$ is the error in observed galaxy spectrum and $N$ is the number of good pixels used in the fitting process.

First, both galaxy and stellar template spectra are rebinned in wavelength space ($x=ln(\lambda)$). A range of model spectra are obtained by convolving the stellar template with a broadening function, which takes into account the true galaxy radial velocity and a range of trial velocity dispersions. The minimisation is then carried out over this range of broadened templates. Compared to the other methods, Cappellari $\&$ Emsellem showed that for low S/N, the use of maximum penalized likelihood suppresses the noise effect in the solution. This makes the pPXF method robust for objects with low S/N such as our dwarf galaxies.

pPXF enables us to use any desired wavelength range for measurements (see Figure~\ref{fig:ppxf} for example). In order to benefit from high S/N part of the spectrum, we chose the rest frame spectral range of 8450-8700 \AA~ which covers the CaT features. All galaxy spectra and stellar templates are continuum subtracted. Due to the noise, for a few faint galaxies ($M_R>-16$ and $S/N<15$), the measured velocity dispersion is very sensitive to the selected wavelength range and therefore, high errors are expected in the final results.

In order to provide realistic initial values of radial velocity to pPXF, the IRAF FXCOR package was used to obtain the relative velocity for each selected pair of the galaxy and template spectra. FXCOR is based on the Fourier cross-correlation technique developed by Tonry \& Davis (1979). The location of the highest peak of the correlation function represents the most probable blue/red-shift value of the galaxy. Although the FXCOR task can measure velocity dispersions, it also requires simulations of multiple spectra in order to construct a proper relation between the width of the highest peak of the correlation function and the real velocity dispersion of the galaxy. Such simulations are not necessary when using pPXF. Other advantages of the pPXF package include the use of multiple stellar templates, where these are optimized and weighted, and shorter computing time.

\begin{figure*}
\centering

\subfigure[] % caption for subfigure a
{
    \label{fig:sub:a}
    \includegraphics[width=3.1in]{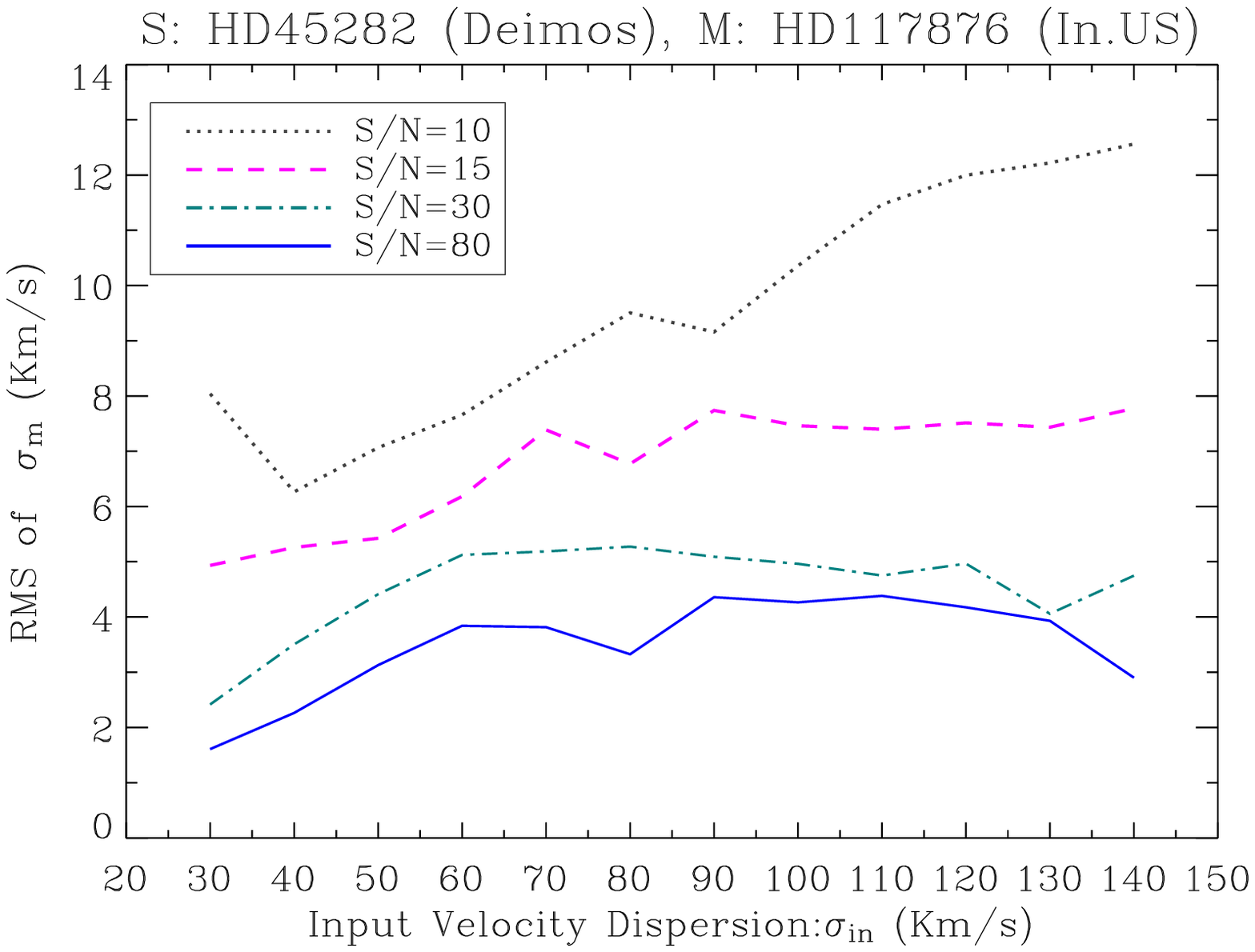}
}
\hspace{-0.1in}
\subfigure[] % caption for subfigure b
{
    \label{fig:sub:b}
    \includegraphics[width=3.1in]{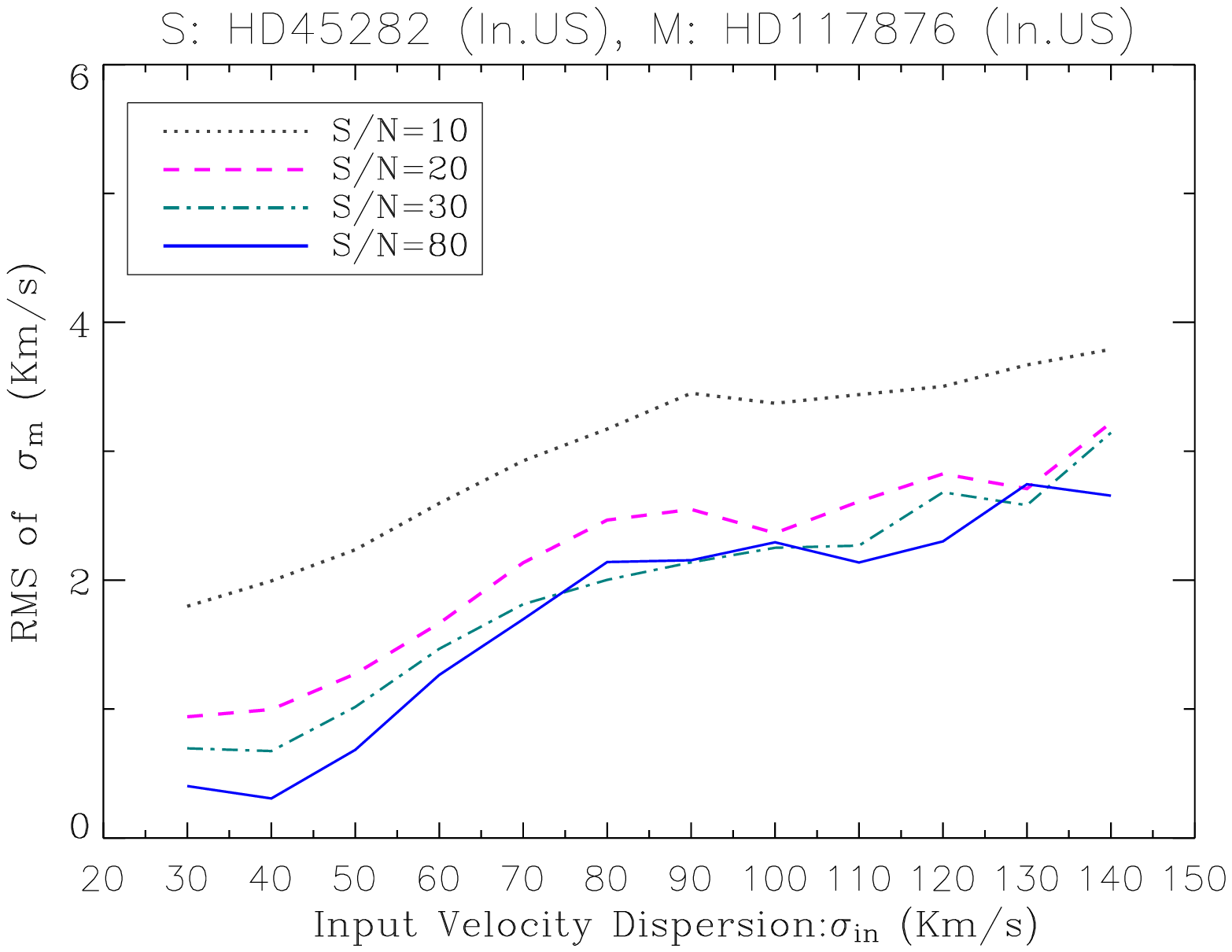}
}
\caption{The RMS of measured velocity dispersion is plotted against the input velocity dispersion ($\sigma_{in}$) used to run simulations. 300 simulated galaxies are generated for each S/N and $\sigma_{in}$. In both plots, the templates are the same. In Plot (a), DEIMOS and Indo-US spectra used as simulation and measurement templates respectively. As one expects, lower S/N and greater $\sigma_{in}$ result in bigger inaccuracies. In plot (b), both templates were selected from Indo-US library. Comparing the 2 plots, it is clear that when using different instruments for the simulation and measurement, the typical value of the uncertainties is greater.}
\label{fig:rms_snr} % caption for the whole figure
\end{figure*}

\begin{table}
\renewcommand{\arraystretch}{0.99}\renewcommand{\tabcolsep}{0.12cm}
\caption{The list of stellar templates, their spectral type and their stellar parameters (i.e. effective temperature, metallicity and surface gravity). T$_{eff}$ and g are in terms of [$^{\circ}K$] and [cm/s$^2$]. All stellar parameters are taken from Indo-US library. These templates were used to perform the initial  measurements and simulations. }
\begin{center}
\begin{tabular}{c | l|l|l}
\hline
Library & \multicolumn{1}{c|}{ID [HD]} & SPTYPE & \footnotesize{T$_{eff}$~~[Fe/H]~~~log(g)} \\ \hline  \hline
DEIMOS & 44007 & G5IV & 4850~~~~-1.71~~~~~~2.00 \\
(This Study) & 45282 & G0V & 5280~~~~-1.52~~~~~~3.12  \\
  \hline
\multirow{9}{*}{Indo-US} & 44007 & G5IV & 4850~~~~-1.71~~~~~~2.00  \\
 & 45282 & G0V & 5280~~~~-1.52~~~~~~3.12  \\
 & 117876 & G8III  & 4582~~~~-0.50~~~~~~2.25  \\
 & 161797 & G5IV  &  5411~~~~~0.16~~~~~~3.87 \\
 & 223-65 & G2IV  & 5350~~~~-0.59~~~~~~3.50  \\
 & 142198 &  K0III   & 4700~~~~-0.08~~~~~~2.99 \\
 & 48433 & K1III  & 4425~~~~-0.24~~~~~~1.35  \\
 & 6497 & K2III  & 4421~~~~~0.02~~~~~~2.80  \\
 & 78479 & K3III  & 4509~~~~~0.57~~~~~~2.54 \\
\hline
\end{tabular}
\label{tab:templates}
\end{center}
\end{table}

\begin{figure*}
\begin{center}
\subfigure[] % caption for subfigure a
{
    \label{fig:sub:a}
    \includegraphics[width=3.2in]{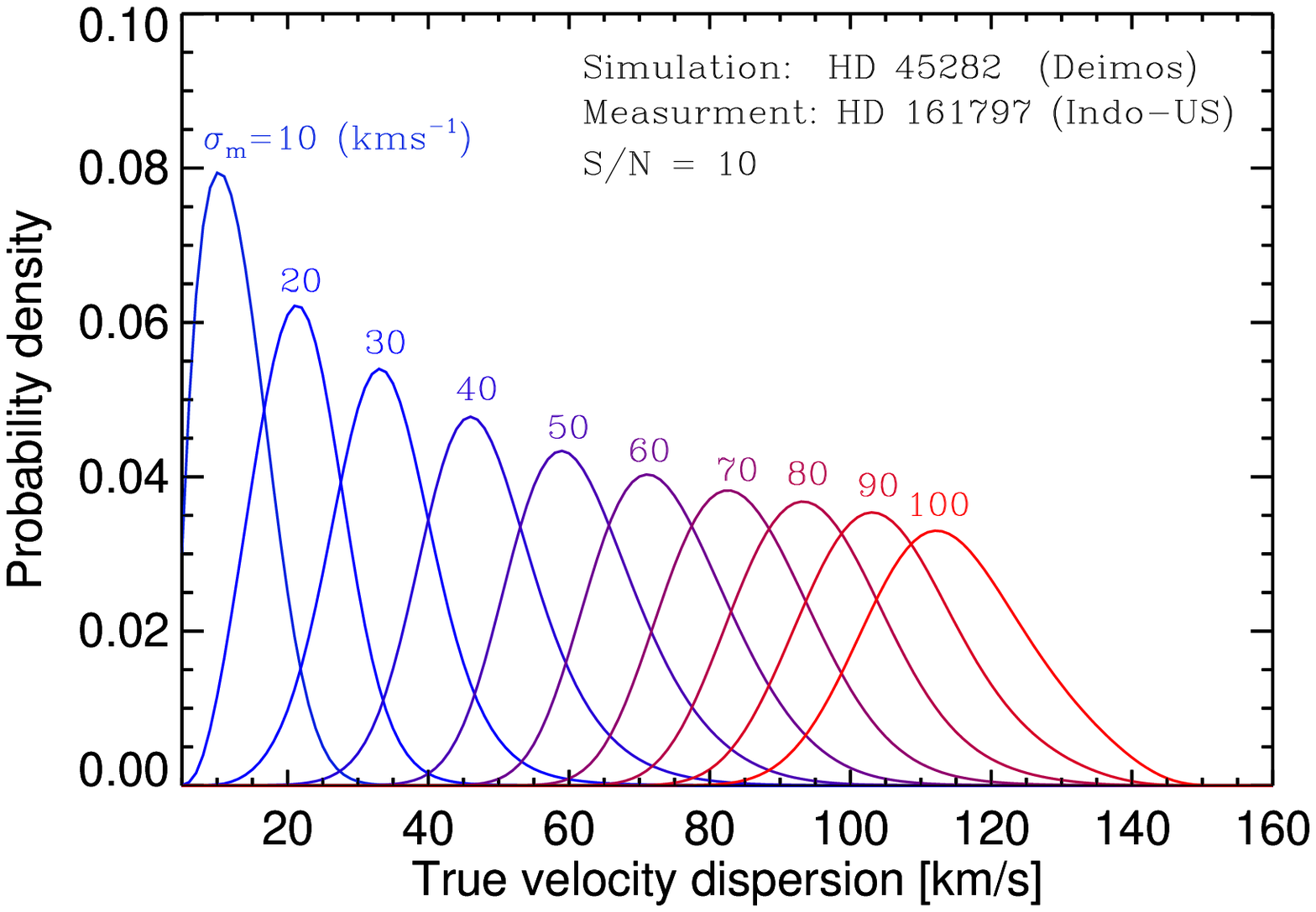}
}
\hspace{-0.30in}
\subfigure[] % caption for subfigure b
{
    \label{fig:sub:b}
    \includegraphics[width=3.2in]{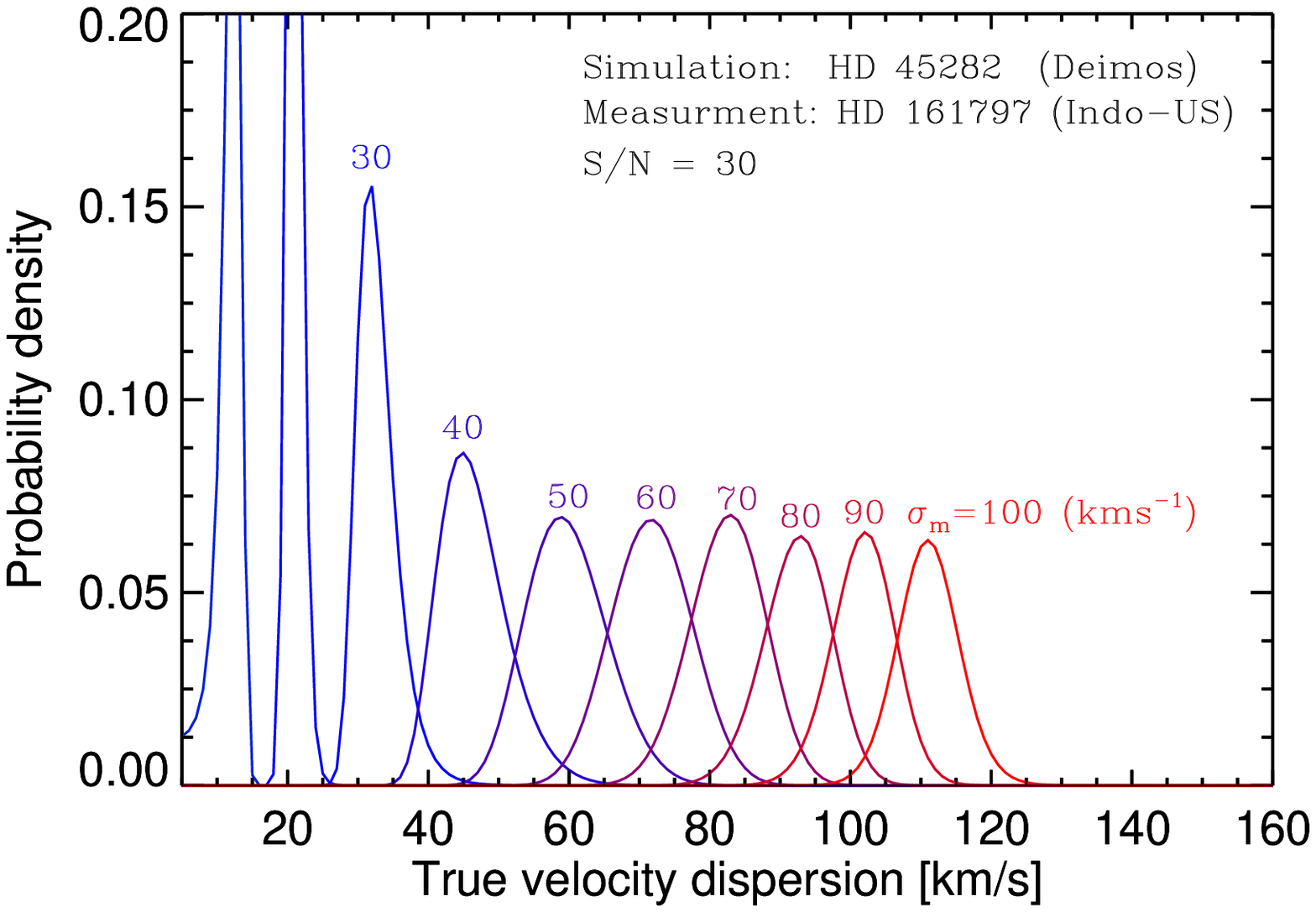}
}
\subfigure[] % caption for subfigure a
{
    \label{fig:sub:a}
    \includegraphics[width=3.2in]{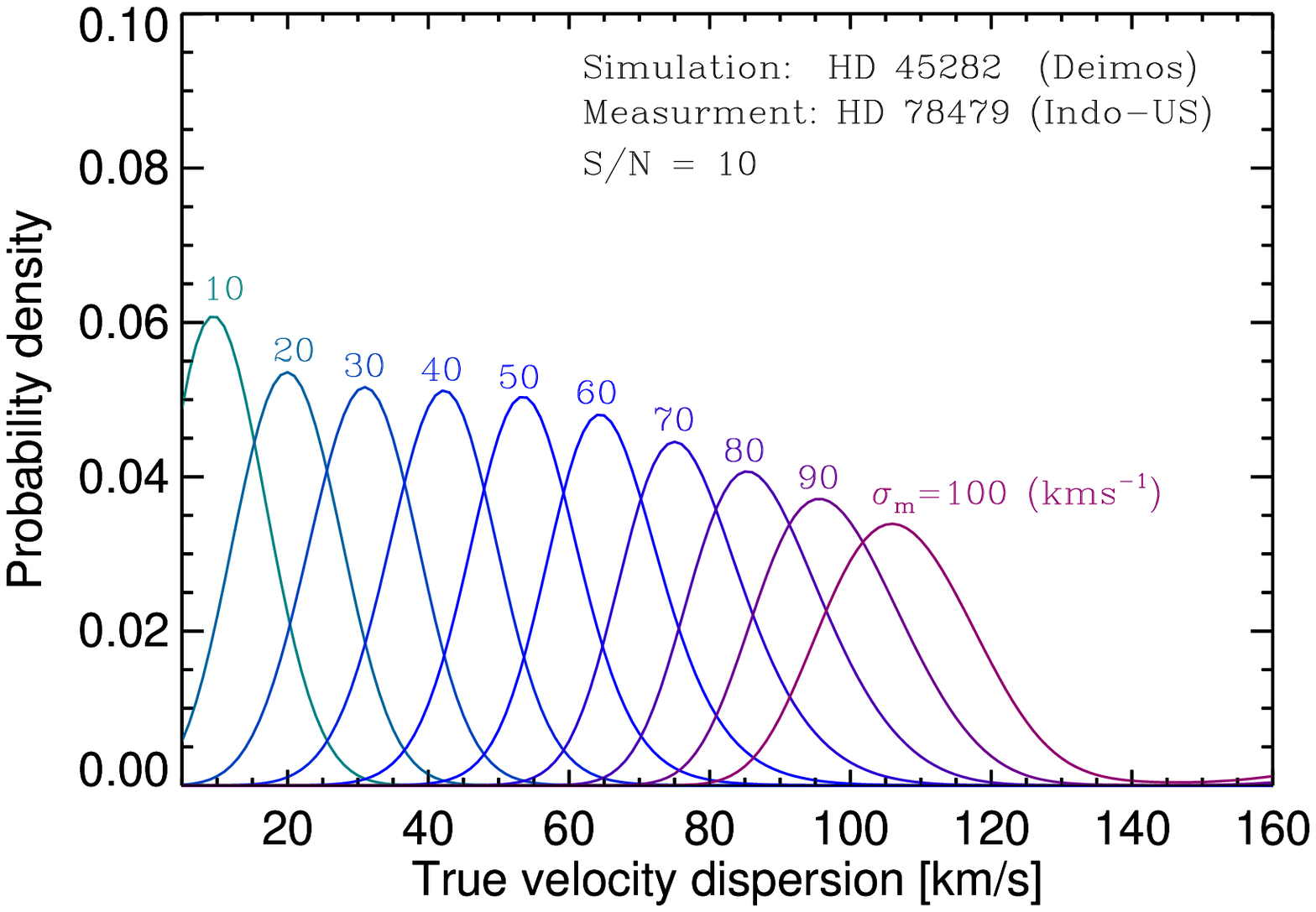}
}
\hspace{-0.30in}
\subfigure[] % caption for subfigure b
{
    \label{fig:sub:b}
    \includegraphics[width=3.2in]{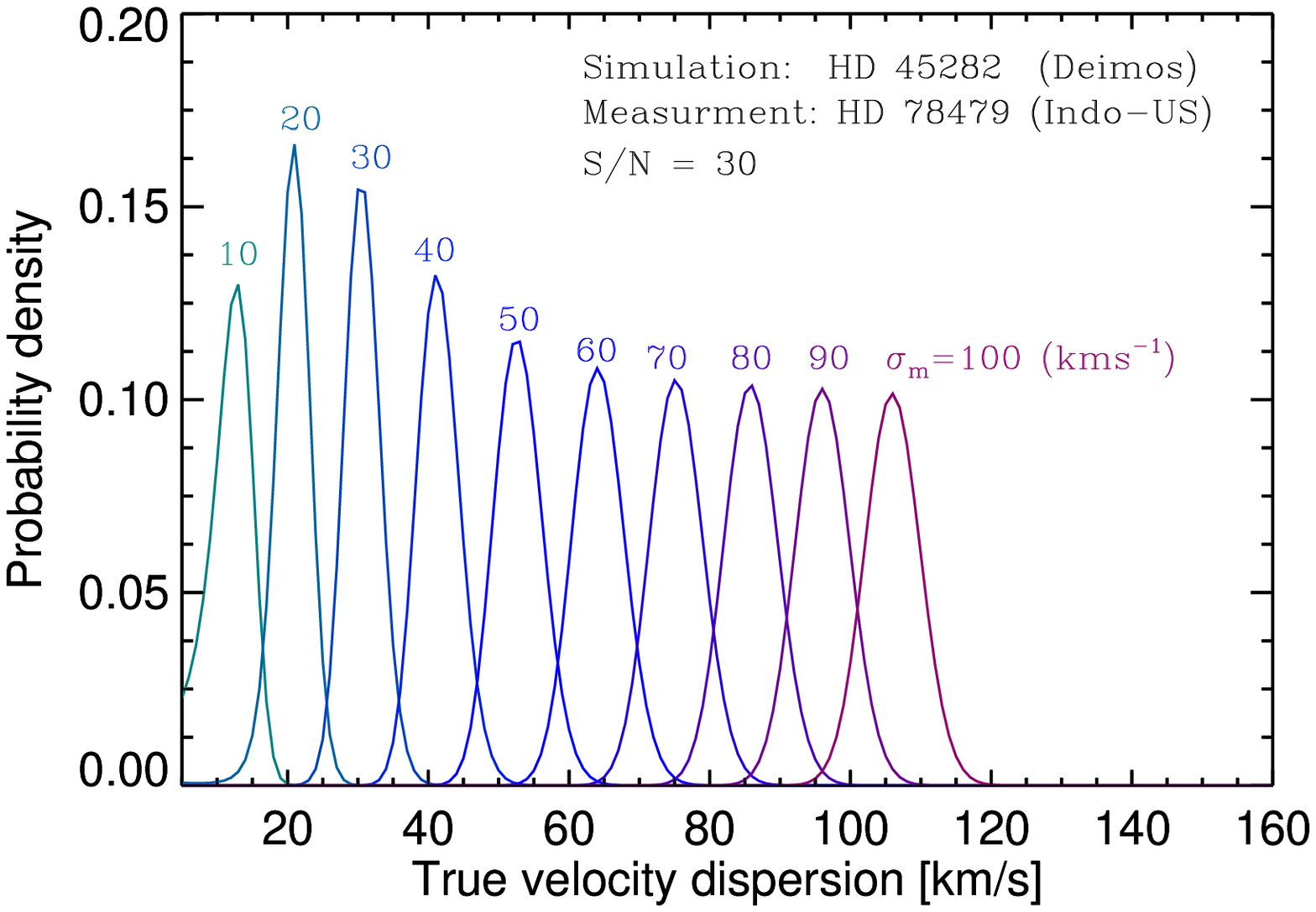}
}

\caption{The distribution of true velocity dispersion for measured dispersion, $\sigma_m$. Different instruments and templates are chosen to consider all sources of uncertainty. For each $\sigma_{in}=10,20,...,100~ km~s^{-1}$, 300 simulated spectra are generated using HD45282. 
In each diagram, the S/N of the simulated spectra is constant. The signal-to-noise ration in left and right panels are 10 and 30, respectively. To see the dependence of the uncertainties upon the stellar template used for measurements, measurements are carried out with HD161797 (G-star) and HD78479 (K-star) in top and bottom panels, respectively and then the statistics are inverted to find $P(\sigma_{true}|\sigma_m)$. The width of the Gaussian peaks corresponds to the statistical uncertainties (see \S \ref{sec:staterror}) and the systematic shift of the peaks along the axis of true velocity dispersion are due to the fact that we used different templates and different instruments for simulation and measurement (see \S \ref{sec:mismatch} for more detailed analysis). We see higher uncertainties for larger $\sigma_m$ values. To estimate the final uncertainty of the measured dispersions, 1$\sigma$ width of each corresponding peak and its systematic shift are added in quadrature.}
\label{fig:errors}
\end{center}
\end{figure*}

Of the eleven stellar templates observed by DEIMOS, only two spectra (i.e. HD44007 and HD45282) had high enough S/N near the strong CaT absorption lines to be used in our analysis. Therefore, we added more G/K-type spectra from the {\it Indo-U.S. Coude Feed} spectral library (Valdes et al, 2004). For practical reasons, especially computing time, we used 11 stellar templates (see Table~\ref{tab:templates}) to perform single template analysis and simulations. To find the best optimal synthetic template, a set of 80 mixed templates were used to finalize the results.

\begin{table}
\renewcommand{\arraystretch}{0.99}\renewcommand{\tabcolsep}{0.12cm}
\caption{The list of most frequent stellar templates used in the analysis with Set50 and Set80. These templates are used to reproduce more than two galaxy spectra. Columns 3, 4 and 5 show the effective temperature, metallicity and surface gravity of each star respectively. Stellar parameters are taken from Indo-US library. The list is sorted in terms of the metallicity. Columns 6 and 7 indicate the utilization frequency of the templates of Set50 and Set80. Set50 does not contain 7 of the templates available in Set80. }
\begin{center}
\begin{tabular}{l c l c c }
\hline
~~ID  & SPTYPE & \footnotesize{T$_{eff}$~~~~[Fe/H]~~~~log(g)} & S50 & S80 \\ 
$[$HD$]$  &  & ~[$K$]~~~~~~~~~~~~~~~~~~~[cm/s$^2$] &    (\%) & (\%)  \\
\hline \hline
63791  &  G0I  &  4700~~~~~-1.81~~~~~~~1.70  &  17  &  10    \\
44007  &  G5IV  &  4850~~~~~-1.71~~~~~~~2.00  &  5  &  7    \\
187111  &  G8IV  &  4429~~~~~-1.54~~~~~~~1.15  &  24  &  22    \\
45282  &  G0V  &  5280~~~~~-1.52~~~~~~~3.12  &  24  &  29    \\
81192  &  G8III  &  4582~~~~~-0.70~~~~~~~2.75  &  7  &  2    \\
210752  &  G0V  &  5910~~~~~-0.64~~~~~~~4.25  &  12  &  20    \\
G\_241-18  &  G5V  &  5511~~~~~-0.61~~~~~~~4.00  &  -  &  10    \\
G\_165-11  &  G0V  &  5725~~~~~-0.56~~~~~~~4.00  &  20  &  17    \\
201099  &  G0V  &  5872~~~~~-0.50~~~~~~~4.06  &  -  &  7    \\
157214  &  G0V  &  5676~~~~~-0.41~~~~~~~4.33  &  -  &  7    \\
99167  &  K5III  &  3930~~~~~-0.38~~~~~~~1.61  &  10  &  10    \\
82210  &  G5III  &  5250~~~~~-0.34~~~~~~~3.42  &  7  &  5    \\
83787  &  K5III  &  4000~~~~~-0.18~~~~~~~1.60  &  17  &  22    \\
169414  &  K2III  &  4450~~~~~-0.16~~~~~~~2.67  &  7  &  7    \\
186486  &  G8III  &  4980~~~~~-0.11~~~~~~~3.08  &  12  &  15    \\
225212  &  -  &  3700~~~~~-0.03~~~~~~~0.80  &  10  &  2    \\
150680  &  G0IV  &  5825~~~~~~0.00~~~~~~~3.80  &  -  &  7    \\
237903  &  -  &  4070~~~~~~0.00~~~~~~~4.70  &  46  &  20    \\
17925  &  K2V  &  5091~~~~~~0.10~~~~~~~4.60  &  10  &  7    \\
161797  &  G5IV  &  5411~~~~~~0.16~~~~~~~3.87  &  10  &  10    \\
63302  &  K3I  &  4500~~~~~~0.17~~~~~~~0.20  &  -  &  10    \\
182293  &  K3IV  &  4486~~~~~~0.25~~~~~~~3.00  &  20  &  17    \\
G\_196-9  &  K5V  &  4000~~~~~~0.28~~~~~~~4.50  &  -  &  24    \\
130705  &  K4III  &  4350~~~~~~0.51~~~~~~~2.10  &  22  &  20    \\
\hline
\label{tab:ftemplates}
\end{tabular}
\end{center}
\end{table}

The first pPXF run shows a significant difference in the velocity dispersion values when different templates in Table~\ref{tab:templates} are used. The scatter in $\sigma$ values gives an estimate of the uncertainty due to the multistellarity of galaxies. To reduce the effect of stellar population on the measured $sigma$, a set of 50 mixed stellar templates (hereafter, Set50) from Indo-US library consisting of G/K-stars were simultaneously given to pPXF and the best combination was obtained by optimizing the $\chi^2$ values. For each galaxy spectrum, the Set50 pPXF run significantly improves upon a single template, the $\chi^2$ values are improved by 10\% (on average), compared with 7\% intrinsic statistical $\chi^2$ deviation of a single template pPXF run. This also points at the multistellar nature of galaxy spectra. The impact of the stellar template mismatch is discussed in \S \ref{sec:mismatch}. In 44\% (20\%) of the cases, the $\chi^2$ is improved by more than 10\% (20\%).

\begin{figure*}
\centering
\subfigure[] % caption for subfigure a
{
    \label{fig:sub:a}
    \includegraphics[width=3.1in]{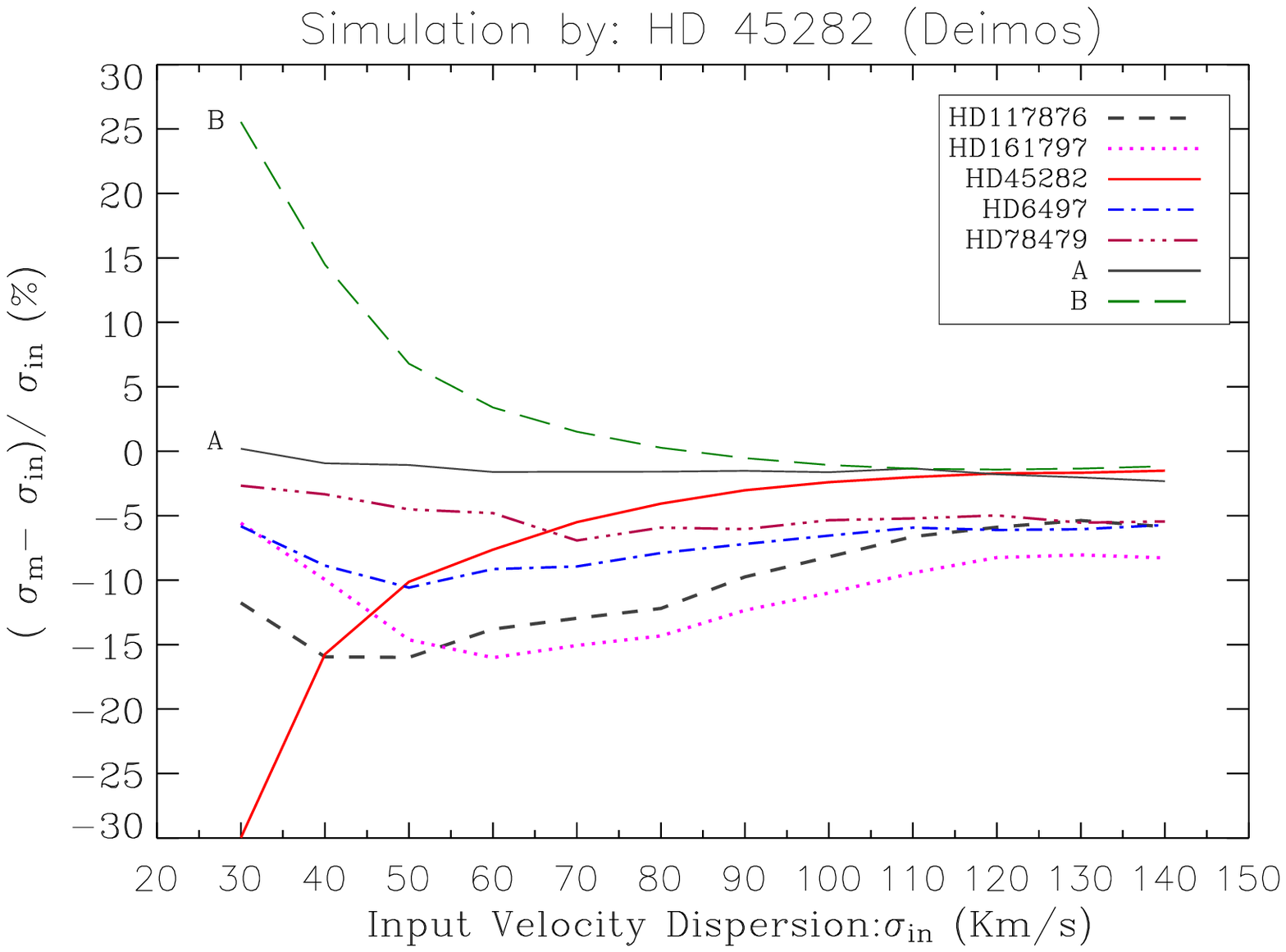}
}
\hspace{-0.1in}
\subfigure[] % caption for subfigure b
{
    \label{fig:sub:b}
    \includegraphics[width=3.1in]{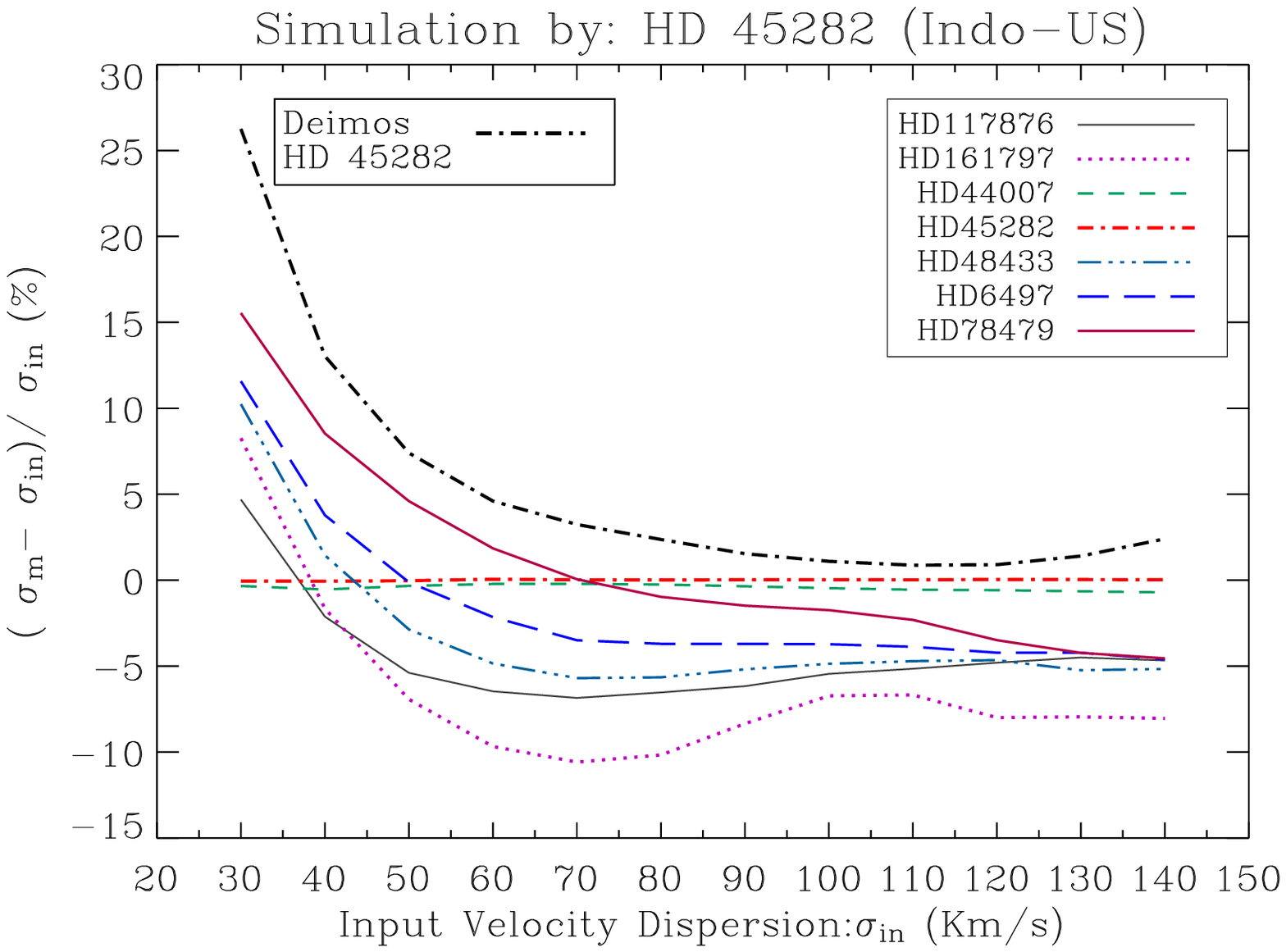}
}
\caption{The relative difference between the measured velocity dispersion and the given input value is plotted for different templates. In panel (a), DEIMOS spectrum HD45282 is used for simulation, and measurements were carried out with Indo-US and DEIMOS spectra. The template spectrum is broadened with the desired broadening factor ($\sigma_{in}$) and no noise is added in this stage to  see only the effect of template mismatch. Curve A shows the results of the same template used for simulation and measurement (i.e. DEIMOS HD45282) and as seen, the discrepancies are negligible. The measurements using another DEIMOS template (i.e. HD44007) leads us to curve B with larger discrepancies for low dispersions. Other curves show the measurement results using Indo-US templates. Panel (b) is the same as panel (a) except for the instrument used for simulation. As expected, there is no discrepancy when the same template is used for measurement (i.e. Indo-US HD45282). The black dash-dotted line in right panel shows the results of measurements with Deimos template (HD45282). As a result, the effect of mismatch and other uncertainties are less than 15\% for $\sigma_{in}>40~km~s^{-1}$. Ignoring both templates HD44007 and HD45282 whose results behave differently regardless of the selected instrument, the average mismatch error is about 7\%. We noted that the metallicity of these two templates are significantly less than the other templates (see Table \ref{tab:templates}).}
\label{fig:discrepant} % caption for the whole figure
\end{figure*}

\begin{figure*}
\centering
\subfigure[] % caption for subfigure a
{
    \label{fig:sub:a}
    \includegraphics[width=3.1in]{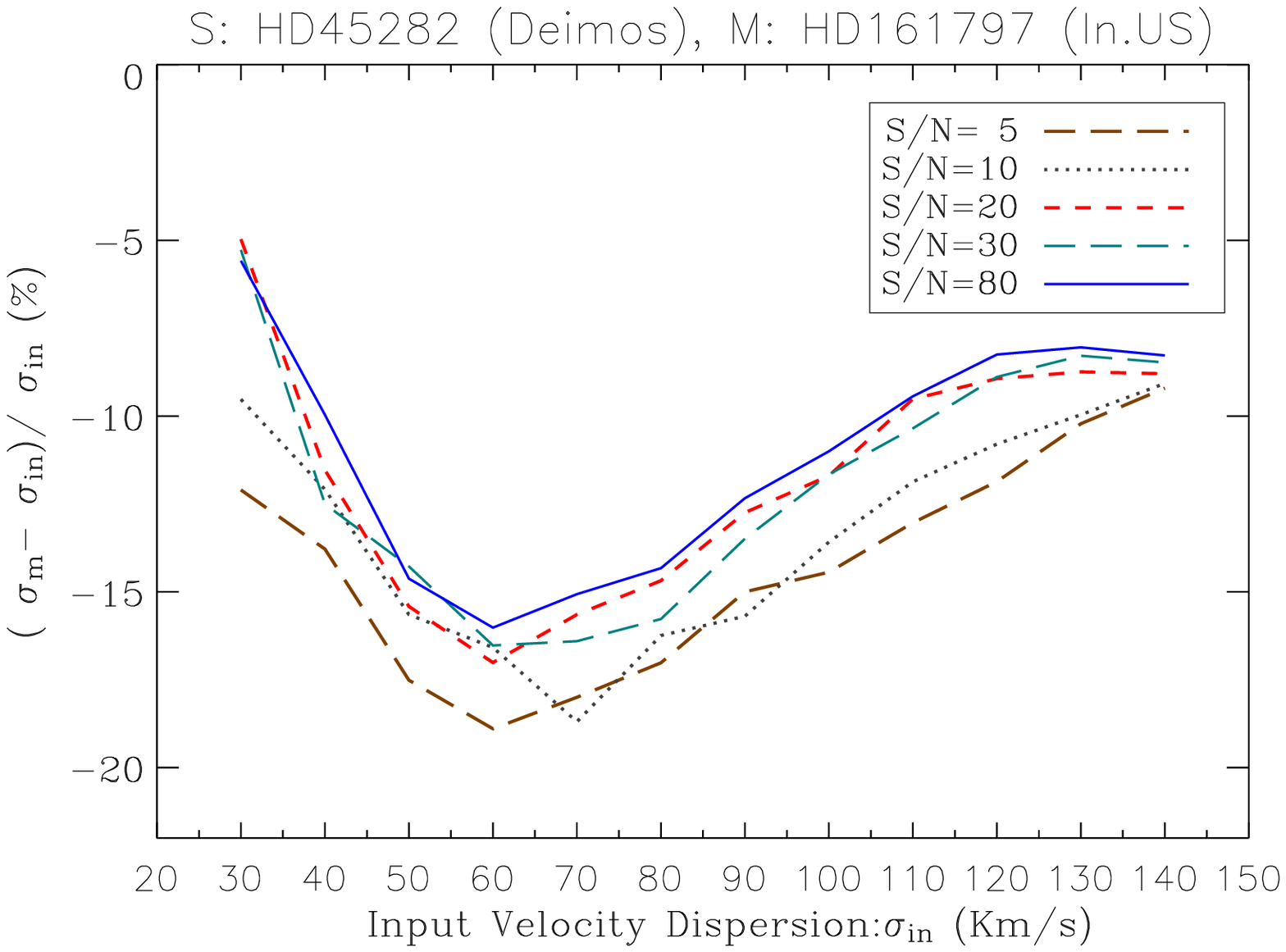}
}
\hspace{-0.1in}
\subfigure[] % caption for subfigure b
{
    \label{fig:sub:b}
    \includegraphics[width=3.1in]{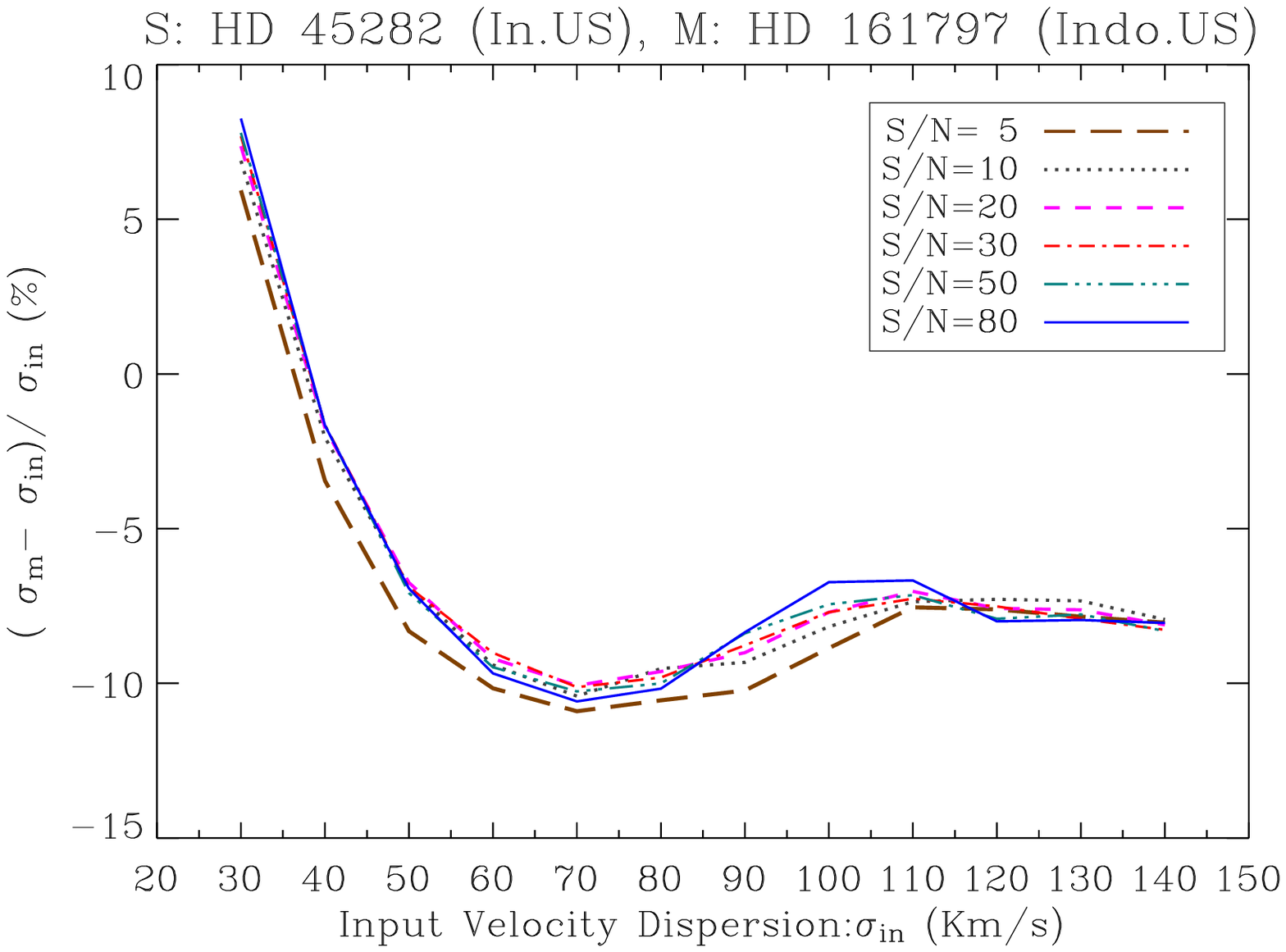}
}
\caption{Same as the Figure ~\ref{fig:discrepant} except that we used only one of the stellar spectra for velocity dispersion measurements and the S/N ratios of the simulated spectra vary. The relative differences of measured and input velocity dispersions are plotted for different S/N values. This shows that for $S/N>10$ per pixel, mismatch and other uncertainties are not affected by signal level which means that the effect of S/N enters only through the statistical uncertainties. 300 simulated spectra were generated for each pair of S/N and $\sigma_{in}$. The simulations for left and right panels were carried out with DEIMOS and Indo-US spectra of HD45282.  The same template (HD161797) was used for measurements in both panels. }
\label{fig:discrepanter} % caption for the whole figure
\end{figure*}

To explore the sensitivity of the final results to the number of templates, a set of 80 templates (hereafter Set80), consisting of Set50 and 30 more G/K-type templates, also chosen from Indo-US library, were used. Comparing the resulting velocity dispersion for Set50 and Set80 indicates that for 36 of 41 galaxies, the discrepancy in the results is less than 2\% and it does not exceed 10\% for the rest. Although a larger number of templates improves the results, it dramatically increases the computation time. Moreover, the use of Set80 instead of Set50 does not improve the $\chi^2$ significantly (i.e. $\sim$1\% on average), and thus we did not continue adding more templates to Set80. We find that all galaxies are modelled with a combination of at least two templates. For 80\% of the galaxies in our sample, more than two templates describe the spectra and in 30\% of the cases, a mixture of more than 4 templates were chosen. The results of the Set80 pPXF run are assigned to each galaxy and used for the rest of this paper (See Table~\ref{tab:results}).

When we used Set80 (Set50), 38 (14) templates are not chosen to describe any of sample galaxies. 6 of 26 used templates from Set50  are not chosen again  by pPXF when using Set80. Of 30 added templates to Set50, only 14 templates contribute to describe at least one galaxy spectrum in Set80 pPXF run. In table \ref{tab:ftemplates}, a list of stellar templates which are used to reproduce more than two galaxy spectra is provided.

%-----------------------------------------------------------------------------
\section{Errors and Uncertainties}
\label{errors}

In this section we discuss the accuracy of our measurements and the sources of uncertainties.

\begin{figure}
\centering
\vspace{-0.3 cm}
\includegraphics[width=7.9cm]{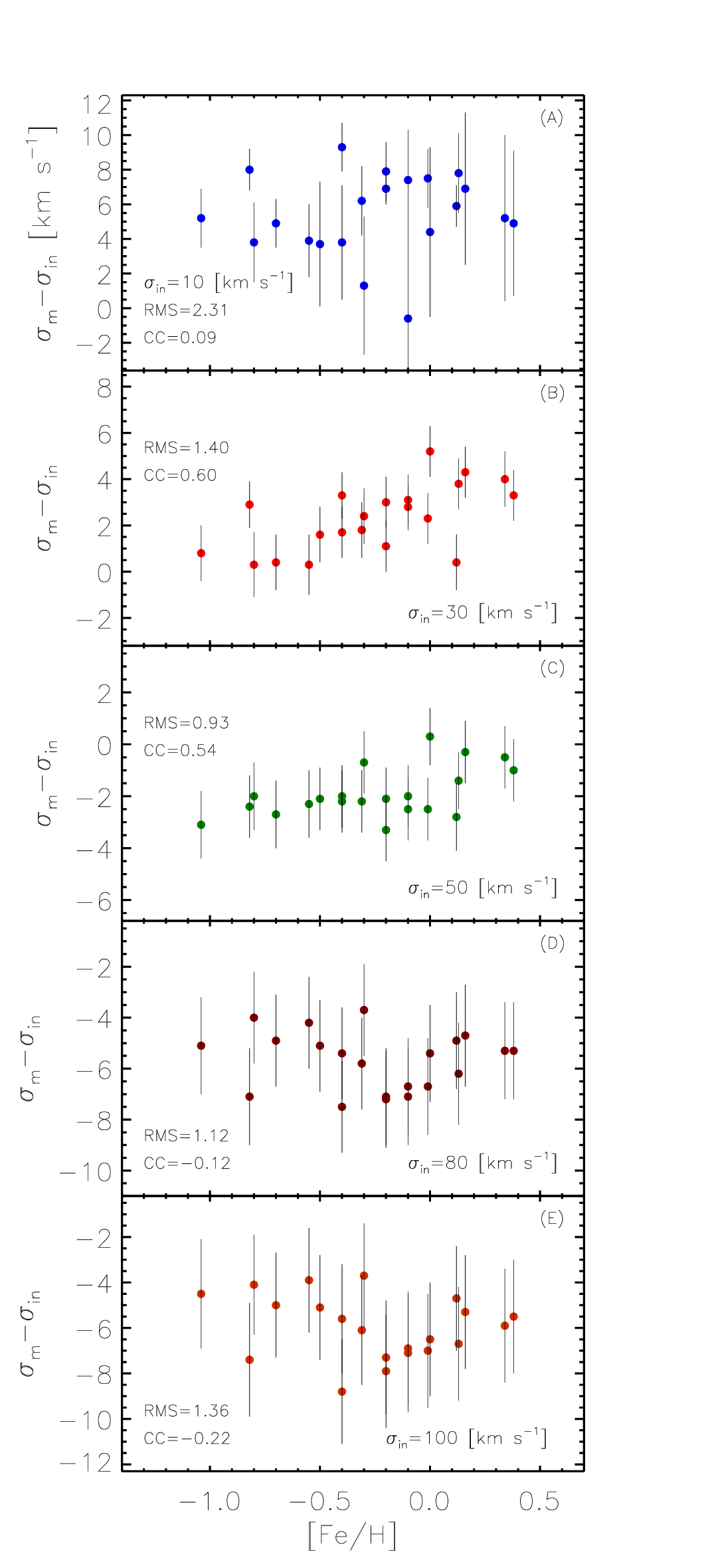}
\caption{The difference between the measured velocity dispersion, $\sigma_{m}$, and the given input velocity dispersion, $\sigma_{in}$, against the metallicity, [Fe/H], of the templates used for measurements. We used the star spectrum HD45282, observed by DEIMOS, as the input spectrum. In each panel, this spectrum was broadened according to the a range of input velocity dispersions. For all simulated spectra, S/N=30. In each case, the error bar is calculated based on 300 simulations. For measurements, we used several templates from Indo-US library with the same spectral type (i.e. G8III) and metallicities ranging between -1 and 0.4 (in solar units). In each panel, RMS is the root-mean-square of the difference between the input and the measured velocity dispersions, and CC shows the coefficient of the correlation between $\sigma_m-\sigma_{in}$ and the metallicity of the templates used for measurements. As seen, the measured velocity dispersions are independent of the metallicity within $\sim 2~ km~ s^{-1}$ uncertainty for $\sigma_{in}> 10~ km ~s^{-1}$.}
\label{fig:simul.set80} % caption for the whole figure
\end{figure}

%\begin{enumerate}
\subsection{Statistical Uncertainties} 
\label{sec:staterror}

One important source of uncertainty is the statistical error which is highly correlated with the S/N. We adopt a ``bootstrapping'' technique to quantify this uncertainty. We simulate an ensemble of galaxy spectra covering the observed range of S/N, the broadening parameter, $(\sigma)$, and the velocity shift $(V_r)$, using the stellar template spectra. To achieve a desired $(\sigma)$, the stellar template spectrum is convolved with a Gaussian function. Then random noise is added to achieve the desired S/N. For each observed galaxy spectra, a set of 300 realizations were generated, noting that, increasing the number of realizations does not alter the results. We then apply the same measurement technique. 

The measurements on each simulated set are carried out with the same initial parameters used in the measurement of original spectrum. Although the input spectral broadening, $\sigma_{in}$, for each set of simulated spectra is the same, the measured value ($\sigma_m$) can be different, due to the noise, and forms a Gaussian distribution around a mean value. The mean of the measured values is not necessarily the same as the input $\sigma_{in}$ value, due to a systematic shift. In general, for all templates in Table~\ref{tab:templates} and all probable S/N values, we obtained $P(\sigma_m|\sigma_{in})$ (i.e. the probability of obtaining $\sigma_m$ for simulated galaxy with the initial velocity dispersion of $\sigma_{in}$). To perform the simulations, for any given signal-to-noise ratio, $\sigma_{in}$ is chosen to have a range of discrete values 10-100 $km~s^{-1}$. For other values of $\sigma_{in}$, we used the polynomial functions to obtain the mean and RMS of the measured velocity dispersion. 
This is explained in detail in section 3.3 of Co09. Figure~\ref{fig:rms_snr} shows the relation between the input velocity dispersion and the RMS of the measured $\sigma$ for different S/N values. As expected, $\sigma_m$ is more uncertain when the S/N decreases or velocity dispersion increases. Left panel of Figure~\ref{fig:rms_snr} indicates that the uncertainty in $\sigma$ is higher when the template for the velocity dispersion measurement is from a different instrument compared to that for the simulated galaxy spectra.

In reality, we have only one spectrum of each galaxy based on which we estimate its velocity dispersion, $\sigma_m$. Similarly, assume that of each set of simulated spectra with given S/N ratio and $\sigma_{in}$, we only have access to one spectrum and its measured velocity dispersion, $\sigma_m$. Therefore, to derive the initial velocity dispersion ($\sigma_{in}$) and its uncertainty, we have to inverse the above statistics and find $P(\sigma_{in}|\sigma_{m})$. This is what needs to be done for any given original galaxy spectrum. Having all $P(\sigma_m|\sigma_{in,i})$ for each $\sigma_{in,i} \in \{1,2,3,...,200~km~s^{-1}\}$, the probability of finding the true velocity dispersion, $\sigma_{true}$ for any measured velocity dispersion, $\sigma_m$, is given by (Co09)

\begin{equation}
P(\sigma_{true,k}|\sigma_{m})=P(\sigma_m|\sigma_{in,k})/\displaystyle\sum_{i=1}^{200}P(\sigma_{m}|\sigma_{in,i}),
\end{equation}

where $\sigma_{in,i}=\sigma_{true,i}=i~km~s^{-1}$. Figure~\ref{fig:errors}, shows 4 examples of the probability density of true velocity dispersion for any given measured value, $\sigma_m$. In order to illustrate the dependency of the results upon the selected templates, for panels ``a" and ``b" of Figure~\ref{fig:errors}, we used the stellar template HD161797 (G-star) and for panels ``c" and ``d" we used HD78479 (K-star). As seen, the peak centre of the probability distributions are not centred upon the true velocity dispersion. This systematic shift is mainly due to the template mismatch and other systematics caused by using different instruments for simulation and measurement. As seen in Figure~\ref{fig:errors}, the location of the peak centres changes when we use different templates for measurements and for a given template it is independent of S/N. For any given velocity dispersion and stellar template, the S/N value affects mainly the statistical uncertainty and therefore the width of the Gaussian peaks. 

To estimate the statistical inaccuracy of the final results, $1\sigma$ uncertainty is derived for each peak of the probability density. The simulated galaxy sample is generated based on the same templates used for the measurements. We performed the above procedure for all individual stellar templates in Table~\ref{tab:templates} which are initially used to measure $\sigma$ and $V_r$ of the sample galaxies. Since all sample galaxies are observed with DEIMOS, in another run of simulations we applied DEIMOS stellar templates (i.e HD44007 or HD45282) for simulation and Indo-US templates for measurement. This method makes our analysis more realistic and enables us to study any additional sources of error due to the instrumental differences. When we apply individual stellar templates, the change in measured velocity dispersions for each galaxy is larger than the corresponding statistical error and we attribute this to the effect of the template mismatch which is investigated in \S \ref{sec:mismatch}.

\subsection{Template Mismatch and Other Source of Uncertainties } 
\label{sec:mismatch}

The velocity dispersion measurement is highly sensitive to the width of the spectral lines in the stellar template based on which the measurements are carried out. The width of the lines is determined by chemical compositions, temperature and the age of the stars (Schulz et al. 2002; Lejeune at al. 1997 and references therein). On the other hand, any galaxy is a composite of different stellar populations and thus lack of any prior knowledge about the exact dominant stellar type is an additional source of uncertainty.

Velocity dispersion measurements of galaxies using the red part of the spectrum are reported to be less sensitive to systematic errors compared with measurements using the blue and visual spectral regions (Pritchet 1978). Our measurements are based on CaT lines which depend weakly on metallicity in metal rich stars (Mallik 1994). However, for metal poor stars, the effect of the metallicity on CaT lines could not be neglected (Starkenburg et al. 2010; Jorgensen et al. 1992; Zhou 1991; Smith \& Drake 1990; D\'iaz et al. 1989) and therefore CaT lines are also employed as a metallicity indicator (Foster et al. 2010; Carrera et al. 2007; Michielsen et al. 2007). Moreover, evidence has been found for significant anti-correlation between Ca II triplet indices and central velocity dispersion of elliptical galaxies (Cenarro et al. 2003). Therefore, template mismatch may have a significant impact on the final results of $\sigma$ measurements and, although, our measurements are based on 80 G/K-type stellar templates, we still need to quantify the uncertainty due to the template mismatch.

Another source of uncertainty appears when the galaxy and template spectra are observed with different instruments. This error is due to the difference in the sensitivity and resolution of the instruments. The full width at half maximum (FWHM) spectral resolution of the Indo-US templates is 1.0 \AA~ and that of DEIMOS is about 1.6 \AA~ for the wavelength range in this study. To take this into account, a proper Gaussian function is convolved with the higher resolution spectrum corresponding to the difference in the spectral resolution. In addition, the sampling rate of Indo-US and DEIMOS spectra are 0.4 \AA/pixel and 0.3 \AA/pixel, respectively. Although pPXF re-samples and re-bins both template and target galaxy spectra to account for this difference, this still may lead to systematic errors in the final results. 

Due to the limited number of stellar templates observed with DEIMOS, we are unable to disentangle the uncertainty due to instrumental differences from the uncertainty due to template mismatch. As seen in Figure~\ref{fig:errors}, the cumulative effect of both template mismatch and difference in the instruments (Indo-US and DEIMOS characteristics) causes systematics in the measured values. Therefore, the shifts seen between the $\sigma_{in}$ and $\sigma_m$ is due to both effects. Similar diagrams are generated for any selected Indo-US and DEIMOS stellar templates. All pairs of different templates and instruments are considered to study all sources of error and their contributions.

We generated a comprehensive set of simulated data using all stellar templates in Table~\ref{tab:templates} according to the recipe explained in section 4.1. In this method, measurements and simulations are carried out with different templates of the same instrument or with the similar templates from different instruments. The resulting velocity dispersions and given initial values are, then, compared after performing the cross measurements on the simulated data with various noise levels. Two examples are presented in Figure~\ref{fig:discrepant} which show the deviation of the measured values, $\sigma_m$, from the true input values, $\sigma_{in}$, for a given velocity dispersion. All simulations are carried out without adding any noise in order to explore the template mismatch effect. We obtained higher systematic errors by choosing the templates from different instruments. All curves in the left panel of Figure \ref{fig:discrepant} illustrate this fact except for two curves A and B for which DEIMOS templates were used for measurements. 

\begin{figure}
\vspace{-0.6cm}
\centering
\includegraphics[width=7.9cm]{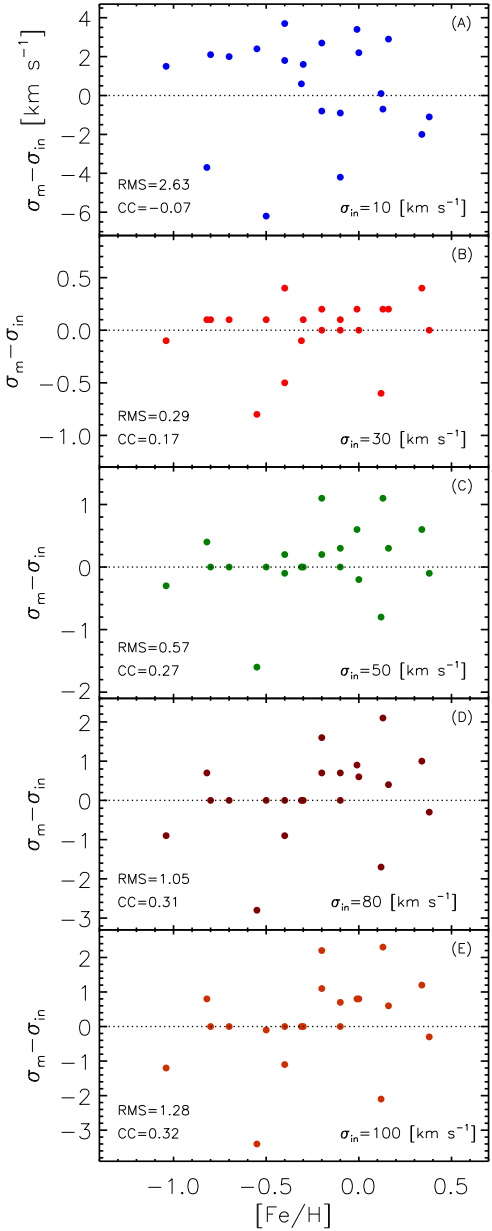}
\caption{Same as the Figure \ref{fig:simul.set80} except the templates used for simulations and measurements. To perform the simulation, we used several templates from Indo-US library with the same spectral type (i.e. G8III) and metallicities ranging between -1 and 0.4. In each panel, the templates were broadened according to the desired input velocity dispersions. In this level of simulations, no noise was added to the simulated spectra. For measurements, we used the Set80 templates. As seen, when using Set80, the measured velocity dispersions are the same as the input velocity dispersions within $\sim 2~ km~ s^{-1}$ uncertainty for $\sigma_{in}> 10~ km ~s^{-1}$. In each panel, RMS is the root-mean-square of the difference between the input and the measured velocity dispersions, and CC shows the coefficient of the correlation between $\sigma_m-\sigma_{in}$ and the metallicity of the templates used for measurements. No significant correlation is seen between the measured and input velocity dispersions when we use Set80. To avoid any confusion, the error bars are not shown in this diagram. Here, the typical uncertainty in measured velocity dispersions is $\sim 2~ km ~s^{-1}$.} 
\label{fig:simul.snr30} % caption for the whole figure
\end{figure}

As expected, no significant systematic shift arises when the same template from the same instrument are used for  simulation and also the measurement. Typically, the relative systematic errors tend to be constant and to slightly increase, respectively, for higher and lower velocity dispersions. To study the sensitivity of the systematics to the S/N, all aforementioned simulations are repeated with random noise added. Left and right panels in Figure~\ref{fig:discrepanter} show the results for the same curves in corresponding panels of the Figure~\ref{fig:discrepant} including different noise levels. As seen, except for very low S/N (i.e. $S/N\lea 10$ per pixel), the systematic errors are not affected by value of S/N. We repeated the procedure several times for different wavelength ranges and any pairs of templates and instruments in Table~\ref{tab:templates}. In all cases, for $\sigma>20~km~s^{-1}$ the effective mean value of all uncertainties does not exceed 20\% of the true value of the velocity dispersion.

Two examples are presented in Tables~\ref{tab:simuldei44007} and \ref{tab:simulin161797}. In Table \ref{tab:simulin161797}, we noticed higher measured $\sigma$ values for HD44007 and HD45282 compared to the measured values using the other templates, even though HD44007 and HD161797 (i.e. the template used for simulation) have the same spectral type. We attribute this to lower metallicity (i.e. [Fe/H]) of both HD44007 and HD45282 compared to the other templates (see \ref{tab:templates}) and consequent systematic errors. As seen in the second and third columns of Table \ref{tab:simulin161797} (low  metallicity stars with [Fe/H]$<-1.5$), for $\sigma_{in}>20~km~s^{-1}$, the average relative discrepancy between the measured values, $\sigma_m$, and input values, $\sigma_{in}$, is 20\%. This value is reduced to 10\% for the other templates. Further evidence for systematic uncertainties in the presence of metallicity differences is seen when measuring the velocity dispersion of our sample galaxies with HD44007 and HD45282 in a single template run which results in 25\% larger values, on average,  compared to those measured using multiple templates. This discrepancy is reduced to $\sim$5\%, on average, for single pPXF run with the other 7 templates of Table \ref{tab:templates}.

\begin{figure}
\begin{center}
\includegraphics[width=8cm]{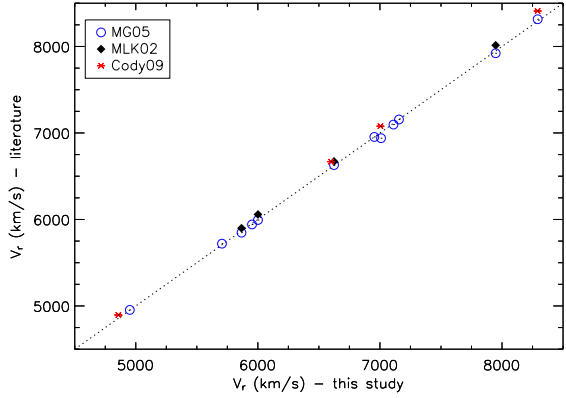}
\caption {Radial velocities from MG05 (open circles) and MLKC02 (filled diamonds) and Co09  (red stars) versus velocities in this study. The dotted line shows equality indicating that the velocities of MLKC02 and Co09 are systematically greater.}
\label{fig:cz}
\end{center}
\end{figure}

\begin{figure*}
\centering
\subfigure[] % caption for subfigure a
{
    \label{fig:sub:a}
    \includegraphics[width=3.1in]{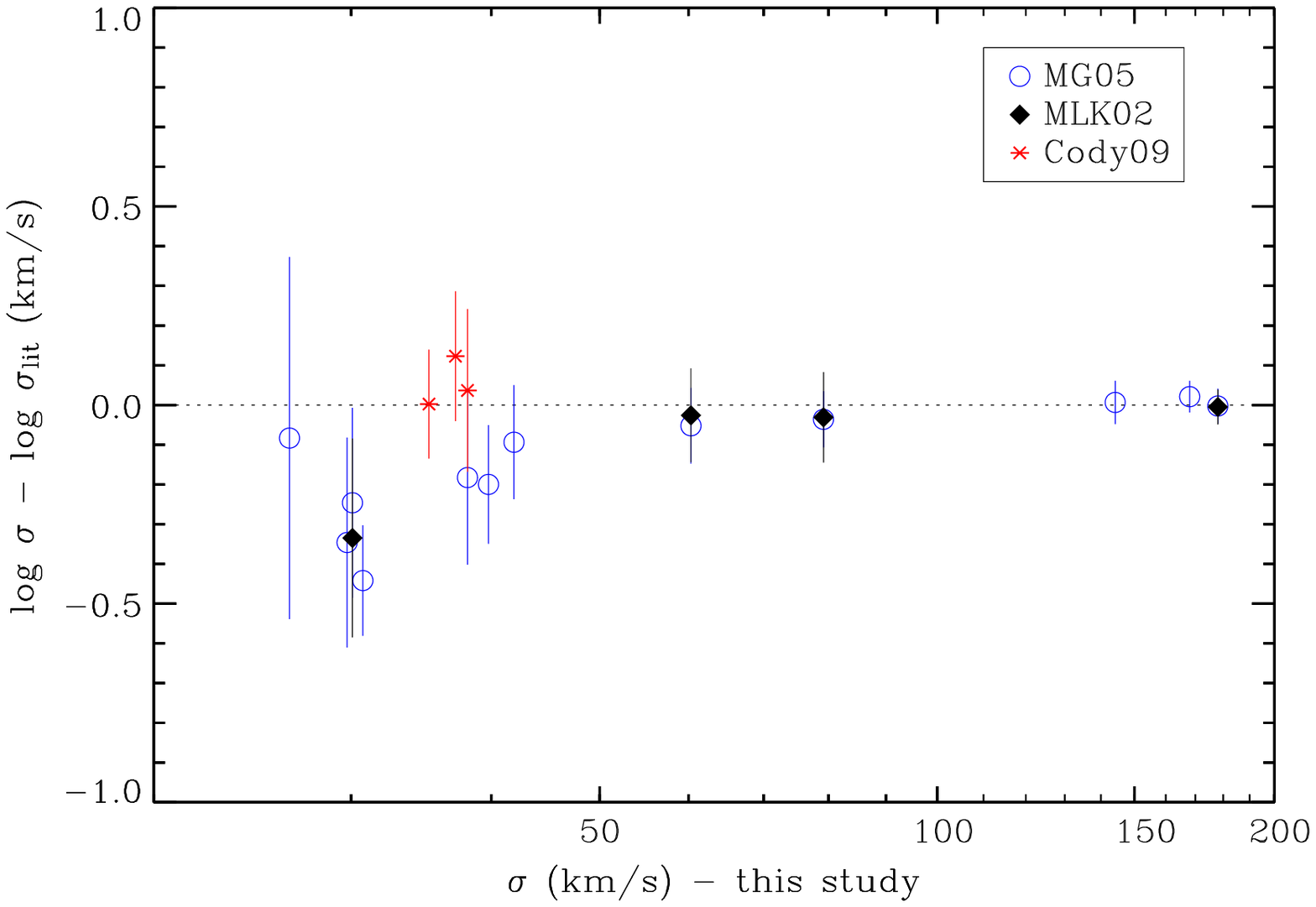}
}
\hspace{-0.1in}
\subfigure[] % caption for subfigure b
{
    \label{fig:sub:b}
    \includegraphics[width=3.1in]{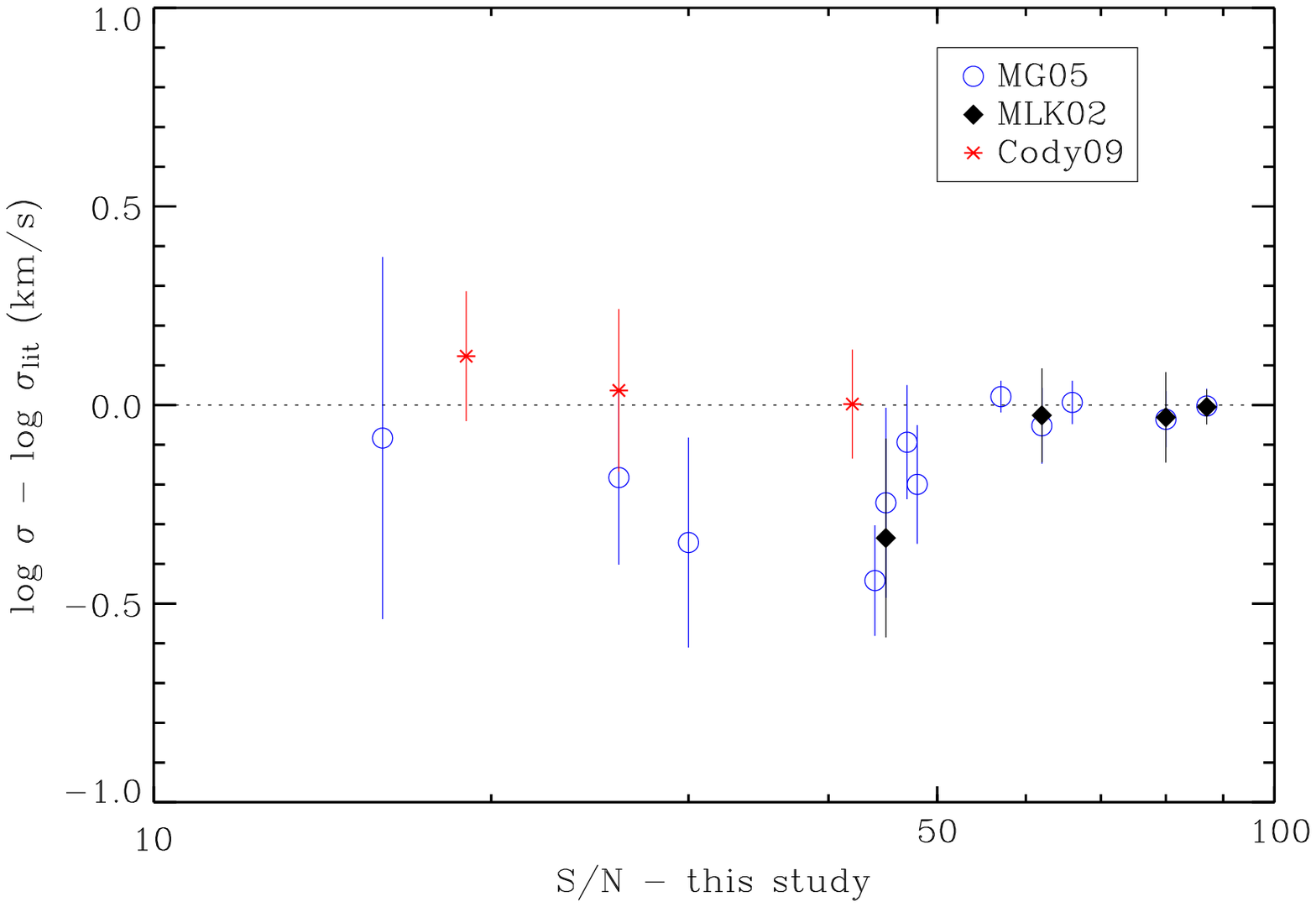}
}
\caption{The difference in measured velocity dispersion in this study and common galaxies in the literature (MG05, MLKC02 and Co09) are plotted against our measured velocity dispersion and S/N value in left and right panel respectively. To derive 1$\sigma$ errors, the errors of our $\sigma$-measurement and those from literature are added in quadrature. The horizontal axis is logarithmic scaled. As seen, except 4 measured values, the rest are consistent with literature within the error bars. There is a good consistency for $\sigma>50~km~s^{-1}$ for which $\langle\sigma/\sigma_{lit}\rangle$ is $0.97\pm0.05$ and is $0.97\pm0.17$ for $\sigma<50~km~s^{-1}$.}
\label{fig:sig} % caption for the whole figure
\end{figure*}

\begin{table*}
%\centering
%\begin{center}
\renewcommand{\arraystretch}{0.9}\renewcommand{\tabcolsep}{0.12cm}
\caption{This table shows the measured velocity dispersion using different stellar templates with different spectral types (see the first row). The first column represents the input velocity dispersion based on which the simulation is carried out. Under each template, the measured values are reported in terms of km s$^{-1}$. DEIMOS star spectrum HD44007 is used to generate the simulated galaxy spectra and all measuring templates are picked up from Indo-US library. For each set, the average of the results is calculated and no noise is added to only estimate the effect of template mismatch which is the offset between the initial and measured values. In the last column, for each row the maximum discrepancy (i.e $\Delta_{max}=|\sigma_m-\sigma_{in}|/\sigma_{in}~(\%)$) is calculated for $\sigma_{in}>10~(km~s^{-1})$, in which $\sigma_{in}$ is the initial velocity dispersion used for the simulation and $\sigma_m$ is the measured value. For $\sigma_{in}>20~(km~s^{-1})$, the average discrepancy is about $7\%$ and for $\sigma_{in}>100~(km~s^{-1})$ it is almost $3\%$. } 
\label{tab:simuldei44007}
\begin{tabular}{| c || c | c | c | c | c | c | c | c | c | c |}
\hline
$\sigma_{in}$ & HD44007 & HD45282 & HD117876 & HD161797 & HD223-65 & HD142198 & HD48433 & HD6497 & HD78479 & $\Delta_{max}$  \\  

(km s$^{-1}$)  & (G5IV) & (G0V) & (G8III) & (G5IV) & (G2IV) & (K0III) & (K1III) & (K2III) & (K3III) &  $(\%)$  \\ \hline \hline

10 & 1 & 11 & 5 & 17 & 14 & 8 & 14 & 12 &  5  & -  \\
20 & 11 & 14 & 20 & 22 & 17 & 20 & 22 & 21 &  22 & 45  \\
30 & 26 & 25 & 27 & 29 & 30 & 27 & 28 & 28 &  29 & 17   \\
40 & 35 & 36 & 34 & 35 & 40 & 34 & 35 & 36 &  39 & 15  \\
50 & 46 & 47 & 43 & 43 & 42 & 43 & 44 & 45 &  49 & 16  \\
60 & 57 & 57 & 52 & 51 & 52 & 53 & 53 & 55 &  58 & 15   \\
70 & 68 & 68 & 62 & 60 & 62 & 62 & 63 & 64 &  67 & 14  \\
80 & 79 & 79 & 72 & 69 & 72 & 72 & 80 & 74 &  76 & 14  \\
90 & 89 & 90 & 82 & 79 & 83 & 83 & 83 & 84 &  86 & 12  \\
100 & 100 & 101 & 93 & 89 & 93 & 100 & 93 & 94 &  100 & 11   \\
110 & 111 & 112 & 104 & 100 & 110 & 104 & 110 & 105 &  110 & 9  \\
120 & 122 & 123 & 120 & 110 & 114 & 115 & 114 & 120 &  116 & 8  \\
130 & 133 & 134 & 125 & 120 & 125 & 126 & 130 & 125 &  126 & 8  \\
140 & 144 & 146 & 136 & 130 & 140 & 136 & 134 & 140 &  136 & 7  \\

\hline
\end{tabular}
%\end{center}
%\captionsetup{margin=40pt,justification=centering,font=Large}
\end{table*}

\begin{table*}
%\centering
\renewcommand{\arraystretch}{0.9}\renewcommand{\tabcolsep}{0.12cm}
\caption{Same as the Table ~\ref{tab:simuldei44007}. The template HD161797 is used for simulation and all templates are selected from Indo-US library. Looking at column 5, as one expects, no discrepancy between the initial and measured value is found. The two last columns ($\Delta_a~\&~\Delta_b$) show the maximum discrepancies as in Table ~\ref{tab:simuldei44007}. Column $\Delta_a$ is calculated for all applied templates and since the templates HD44007 and HD45282 result in large discrepancies, the column $\Delta_b$ is created ignoring them. Considering all templates, for $\sigma_{in}>20~(km~s^{-1})$, the average discrepancy is about $10\%$ and for $\sigma_{in}>100~(km~s^{-1})$ it is almost $7\%$. The metallicity of HD44007 and HD45282 are lower than the other templates and therefore results in measuring larger $\sigma_m$ compared to the other templates (see Table \ref{tab:templates}).}

\label{tab:simulin161797}
\begin{tabular}{| c || c | c | c | c | c | c | c | c | c | c | c |}
\hline
$\sigma_{in}$ & HD44007 & HD45282 & HD117876 & HD161797 & HD223-65 & HD142198 & HD48433 & HD6497 & HD78479 & $\Delta_a$ & $\Delta_b$ \\  
 (km s$^{-1}$) & (G5IV) & (G0V) & (G8III) & (G5IV) & (G2IV) & (K0III) & (K1III) & (K2III) & (K3III) & $(\%)$ & $(\%)$ \\ \hline \hline

10 & 24.7 & 25.6 & 5.9 & 3.8 & 10.9 & 3.4 & 4.0 & 17.4 & 19.4 & - & -\\
20 & 32.3 & 33.2 & 23.5 & 20.0 & 24.2 & 22.8 & 22.2 & 23.4 & 24.8 & 66 & 24 \\
30 & 40.9 & 42.2 & 33.4 & 30.0 & 33.8 & 32.4 & 32.2 & 33.7 & 35.3 & 41 & 18 \\
40 & 51.1 & 52.3 & 43.7 & 40.0 & 44.0 & 43.0 & 42.6 & 44.2 & 46.8 & 31 & 17 \\
50 & 62.4 & 63.7 & 54.3 & 50.0 & 54.7 & 53.7 & 52.9 & 54.7 & 57.5 & 27 & 15 \\
60 & 73.8 & 75.1 & 64.9 & 60.0 & 65.7 & 64.0 & 63.6 & 65.1 & 67.4 & 25 & 12 \\
70 & 84.8 & 86.1 & 74.5 & 70.0 & 76.5 & 74.4 & 73.8 & 75.2 & 77.0 & 23 & 10 \\
80 & 95.4 & 96.8 & 84.7 & 80.0 & 87.0 & 84.4 & 84.0 & 85.5 & 86.8 & 21 & 9 \\
90 & 105.9 & 107.5 & 94.8 & 90.0 & 97.3 & 94.3 & 93.9 & 95.4 & 96.9 & 19 & 8 \\
100 & 116.4 & 118.3 & 105.4 & 100.0 & 107.5 & 104.8 & 103.9 & 105.7 & 106.7 & 18 & 8 \\
110 & 127.0 & 129.1 & 115.8 & 110.0 & 117.7 & 115.2 & 114.1 & 115.8 & 116.7 & 17 & 7 \\
120 & 137.6 & 140.0 & 125.7 & 120.0 & 128.0 & 125.1 & 124.1 & 126.3 & 126.4 & 17 & 7 \\
130 & 148.4 & 150.9 & 136.5 & 130.0 & 138.2 & 135.6 & 134.2 & 136.2 & 136.2 & 16 & 6 \\
140 & 159.2 & 161.9 & 147.1 & 140.0 & 148.5 & 146.2 & 144.0 & 146.6 & 146.1 & 16 & 6 \\

\hline
\end{tabular}
\end{table*}

In order to examine the effect of the galaxy and stellar template metallicity on measured velocity dispersions, we used 21 stellar templates with the same spectral type (i.e. G8III) from Indo-US library and different metallicity. As seen in Figure \ref{fig:simul.set80}, we used the stellar spectra HD45282, observed by DEIMOS, to simulate galaxies with desired velocity dispersions, i.e. $\sigma_{in}= 10,30,50,80,100~km~s^{-1}$. In each case, 300 simulated spectra with S/N=30 were created. Then, the velocity dispersion of the simulated galaxies were measured using the stellar templates with the same spectral type but different metallicity. Figure, \ref{fig:simul.set80} shows that differences between the measured and input velocity dispersion are independent of the template metallicity within $\sim 2~ km~ s^{-1}$ uncertainty in measured velocity dispersions. Due to the template mismatch effect, in each panel of the Figure \ref{fig:simul.set80}, there is an offset between the given velocity dispersion and the mean value of the measured velocity dispersions using the templates with different metallicity. This offset is due to the template mismatch effect and for a given spectral type, it seems to be independent of the template metallicity. 

In Figure \ref{fig:simul.snr30}, we used the mixed stellar templates, i.e. Set80, for measuring the velocity dispersion of simulated galaxies with different metallicities. To perform the simulations, we broadened the stellar templates with different metallicities (i.e. the G8III-type stars used in Figure \ref{fig:simul.set80}) according to desired input velocity dispersions. At very low velocity dispersion (i.e. $\sigma_{in}=10~km~s^{-1}$) the mean value of the measured velocity dispersion for all simulated spectra with different metallicities is $11.5~km~s^{-1}$. For $\sigma> 10~km~s^{-1}$, the measurements result in the same value as the input values and are independent of metallicity. As a conclusion, based on the simulations, using the mixed stellar templates results in more reliable measurements and reduces the effect of template mismatch. To confirm these results, we measured the velocity dispersion of the simulated spectra in all panels of Figure \ref{fig:simul.set80} using Set80. For panels of Figure \ref{fig:simul.set80}, from top to bottom, the results are $\sigma_m=10.5\pm2.3,29.7\pm1.3,49.8\pm1.3,79.5\pm1.9$ and $99.3\pm2.5 ~km~s^{-1}$, respectively. As expected, there is no significant offset from the input velocity dispersion when we use the mixed stellar templates.

\begin{table*}

%\fontsize{10}{12}\selectfont 

\caption{Radial velocity, $V_r$ and velocity dispersion, $\sigma$ of dwarf galaxies in our sample. In columns 10 and 11, we have a list of common galaxies in literature and their $\sigma$. In column 12, the data denoted by asterisks are from MLKC02 and the rests are from Co09.  The likely errors in measured $\sigma$ due to the effect of template mismatch are illustrated in column 9. The statistical uncertainties in $\sigma$ are listed in column 10. Column 8 shows the final uncertainty of the $\sigma$ for all uncertainty sources and is obtained by adding the columns 9 and 10 in quadrature. Column 13 is the S/N of the galaxy spectra which are calculated per pixel (0.33 \AA). Column 14 shows the number of stellar templates used to describe the galaxy spectra in Set80 run.}

\renewcommand{\arraystretch}{0.9}\renewcommand{\tabcolsep}{0.085cm}
\begin{tabular}[tc]{ c  c  c  c  c  c  c  c  c  c  c  c  c c}

%\endhead 

\hline
GMP & RA &  Dec &  $R_{sdss}$ & $cz$ & $\delta cz $ & $\sigma$  &  $\Delta \sigma$   &  $\delta \sigma_1$    &  $\delta \sigma_2$   & $\sigma~(MG05)$ & $\sigma~(Lit.)$  & S/N & \# of  \\  
ID  &  (J2000)  &  (J2000) & mag & $km~s^{-1}$ & $km~s^{-1}$ & $km~s^{-1}$ &   $km~s^{-1}$ & $km~s^{-1}$ & $km~s^{-1}$ & $km~s^{-1}$ & $km~s^{-1}$  & 1/pix & Temp. \\
(1) & (2) & (3) & (4) &  (5) & (6) &  (7) & (8) &  (9) &  (10) &  (11) &  (12) &  (13) &  (14) \\ \hline \hline

3534   &   12:59:21.40   &   27:58:24.80   &   16.13   &   6001   &   7   &   30   &   7   &   7   &   3   &           $53\pm3$          &           $65\pm6^{*}$          &   45   &   7   \\
3474   &   12:59:26.40   &   27:59:18.30   &   18.69   &   7666   &   6   &   22   &   4   &   3   &   2   &                               &                               &   37   &   5   \\
3471   &   12:59:26.60   &   27:59:54.50   &   15.35   &   6624   &   13   &   79   &   5   &   5   &   2   &           $86\pm3$          &           $85\pm8^*$          &   80   &   7   \\
3438   &   12:59:28.50   &   28:01:09.40   &   18.05   &   5952   &   31   &   26   &   8   &   7   &   3   &           $32\pm11$          &                               &   16   &   3   \\
3424   &   12:59:29.30   &   27:56:32.00   &   18.27   &   5596   &   7   &   27   &   9   &   7   &   5   &                               &                               &   29   &   6   \\
3406   &   12:59:30.30   &   28:01:15.10   &   17.41   &   7116   &   7   &   36   &   3   &   2   &   2   &                               &                               &   29   &   3   \\
3376   &   12:59:32.10   &   27:55:15.80   &   17.25   &   7004   &   22   &   37   &   3   &   3   &   2   &                               &           $28\pm4$          &   19   &   5   \\
3340   &   12:59:35.20   &   27:56:05.00   &   17.73   &   4476   &   12   &   43   &   8   &   6   &   5   &                               &                               &   17   &   2   \\
3336   &   12:59:35.50   &   27:54:21.60   &   17.61   &   6954   &   6   &   42   &   4   &   4   &   2   &           $52\pm5$          &                               &   47   &   5   \\
3325   &   12:59:36.00   &   27:54:22.00   &   17.58   &   8953   &   13   &   38   &   7   &   6   &   2   &                               &                               &   18   &   2   \\
3312   &   12:59:37.00   &   28:01:07.00   &   17.42   &   7156   &   27   &   31   &   4   &   3   &   2   &           $85\pm3$          &                               &   44   &   5   \\
3308   &   12:59:37.20   &   27:58:19.90   &   18.09   &   7522   &   8   &   50   &   4   &   3   &   2   &                               &                               &   57   &   4   \\
3296   &   12:59:37.90   &   27:54:26.40   &   14.3   &   7948   &   21   &   178   &   7   &   5   &   4   &           $179\pm4$          &           $180\pm4^*$          &   87   &   1   \\
3292   &   12:59:38.00   &   28:00:03.70   &   16.25   &   4951   &   8   &   40   &   5   &   3   &   3   &           $63\pm6$          &                               &   48   &   5   \\
3223   &   12:59:42.40   &   28:01:58.60   &   18.41   &   7797   &   16   &   22   &   9   &   8   &   5   &                               &                               &   11   &   3   \\
3209   &   12:59:44.20   &   28:00:47.00   &   18.58   &   7111   &   17   &   30   &   5   &   4   &   2   &           $66\pm14$            &                               &   30   &   5   \\
3166   &   12:59:46.90   &   27:59:30.90   &   17.39   &   8292   &   7   &   38   &   6   &   6   &   3   &           $58\pm8$          &           $35\pm4$          &   26   &   4   \\
3146   &   12:59:48.60   &   27:58:58.00   &   17.74   &   5358   &   10   &   38   &   11   &   9   &   6   &                               &                               &   11   &   3   \\
3141   &   12:59:49.10   &   27:58:33.90   &   19.86   &   4957   &   26   &   59   &   15   &   11   &   10   &                               &                               &   5   &   2   \\
3131   &   12:59:50.20   &   27:54:45.50   &   17.29   &   7183   &   14   &   19   &   7   &   6   &   4   &                               &                               &   22   &   4   \\
3119   &   12:59:51.50   &   27:59:35.50   &   19.74   &   7060   &   14   &   37   &   10   &   9   &   5   &                               &                               &   8   &   3   \\
3098   &   12:59:53.90   &   27:58:13.70   &   17.73   &   6740   &   8   &   33   &   7   &   5   &   5   &                               &                               &   28   &   5   \\
3080   &   12:59:55.70   &   27:55:03.80   &   18.11   &   6599   &   10   &   9   &   8   &   6   &   5   &                               &           $16\pm4$          &   19   &   5   \\
3018   &   13:00:01.00   &   27:59:29.60   &   18.17   &   7464   &   10   &   38   &   5   &   5   &   3   &                               &                               &   15   &   3   \\
2983   &   13:00:04.00   &   28:00:30.70   &   19.03   &   6349   &   10   &   27   &   6   &   5   &   3   &                               &                               &   18   &   3   \\
2960   &   13:00:05.40   &   28:01:28.00   &   15.18   &   5866   &   7   &   60   &   5   &   5   &   3   &           $68\pm2$          &           $64\pm5^*$          &   62   &   5   \\
2931   &   13:00:07.10   &   27:55:51.50   &   17.59   &   7704   &   17   &   35   &   6   &   4   &   4   &                               &                               &   47   &   5   \\
2877   &   13:00:11.40   &   27:54:36.40   &   18.71   &   7225   &   12   &   29   &   9   &   8   &   5   &                               &                               &   14   &   5   \\
2839   &   13:00:14.70   &   28:02:26.90   &   14.43   &   5706   &   15   &   168   &   6   &   5   &   3   &           $160\pm3$          &                               &   57   &   4   \\
2808   &   13:00:17.00   &   27:54:16.10   &   19.64   &   8944   &   16   &   69   &   12   &   9   &   8   &                               &                               &   13   &   2   \\
2780   &   13:00:18.70   &   27:55:12.70   &   19.06   &   6554   &   18   &   63   &   12   &   9   &   8   &                               &                               &   13   &   2   \\
2755   &   13:00:20.20   &   27:59:37.60   &   18.87   &   6812   &   13   &   27   &   9   &   8   &   4   &                               &                               &   14   &   2   \\
2736   &   13:00:21.70   &   27:53:54.80   &   16.41   &   4857   &   5   &   35   &   4   &   3   &   2   &                               &           $35\pm3$          &   42   &   4   \\
2718   &   13:00:22.70   &   27:57:55.00   &   18.76   &   6360   &   12   &   30   &   8   &   7   &   4   &                               &                               &   13   &   4   \\
2676   &   13:00:26.20   &   28:00:32.00   &   17.86   &   5516   &   30   &   37   &   8   &   7   &   4   &                               &                               &   19   &   4   \\
2655   &   13:00:27.90   &   27:59:16.50   &   19.46   &   9412   &   14   &   45   &   5   &   5   &   2   &                               &                               &   12   &   3   \\
2654   &   13:00:28.00   &   27:57:21.60   &   14.87   &   7009   &   21   &   144   &   7   &   6   &   3   &           $142\pm4$          &                               &   66   &   3   \\
2605   &   13:00:33.30   &   27:58:49.40   &   18.28   &   5126   &   12   &   36   &   10   &   8   &   6   &                               &                               &   13   &   3   \\
2591   &   13:00:34.40   &   27:56:05.00   &   17.17   &   8653   &   14   &   53   &   9   &   8   &   4   &                               &                               &   20   &   2   \\
2571   &   13:00:36.60   &   27:55:52.20   &   18.68   &   5944   &   18   &   17   &   6   &   5   &   3   &                               &                               &   20   &   3   \\
2563   &   13:00:37.30   &   27:54:41.10   &   19.5   &   5895   &   11   &   25   &   10   &   9   &   5   &                               &                               &   9   &   2   \\

\hline
\end{tabular}
\label{tab:results}
\end{table*}

\subsection{Best-Fit Error Analysis}
\label{sec:bestfiterror}

In addition to the above, we repeated the same analysis for each galaxy to construct an ensemble of simulated spectra by adding the corresponding random noise to the best fits provided by pPXF (e.g red curve in Fig. \ref{fig:ppxf}). This technique effectively gives the statistical errors of both measured radial velocity and velocity dispersion. In the case of having few templates (e.g. when using Set50 or Set80), this method is applied with the use of the equivalent synthetic template based on the best output weights. Applying the multiple stellar templates reduces the effect of the template mismatch particularly when dealing with low S/N spectra of faint galaxies which are expected to have smaller mass and internal velocity dispersion.

\subsection{The Catalogue Uncertainties}
\label{sec:catalogue}

All the stellar templates in Table \ref{tab:templates} were used to measure velocity dispersions for each galaxy in our sample. We used the RMS scatter around the mean values as the likely template mismatch uncertainty (see column 9 of the Table \ref{tab:results}). Since the exact combination of the stellar templates is unknown for each galaxy, the exact value of the template mismatch uncertainty depends on the selected templates for measurements. Even though Table \ref{tab:templates} covers various metallicities and spectral types, choosing another set of templates may slightly alter this estimate, due to the multistellarity of galaxies. For 66\% of the galaxies in Table \ref{tab:results}, the mismatch error is estimated to be less than 20\%  and only exceeds 25\% for 22\% of the galaxies. The average template mismatch errors are ~4\%, ~12\% and  ~18\% for galaxies with the measured velocity dispersions greater than 100 $km~s^{-1}$, between 50 \& 100 $km~s^{-1}$ and between 20 \& 50 $km~s^{-1}$, respectively. These results which are purely obtained from observation are consistent with those estimated using simulations described in \S\ref{sec:mismatch}. Figure \ref{fig:discrepant} shows that the mismatch effect decreases as the velocity dispersion increases. Simulations indicate that the average uncertainty due to the template mismatch is about 15\% for $\sigma\sim50~km~s^{-1}$ and $\sim$5\% for $\sigma\gea100~km~s^{-1}$, which is consistent with the reported values in column 9 of the Table \ref{tab:results}.

In \S \ref{chap:velocities}, it is mentioned that using Set50 improves the $\chi^2$ by $\sim$10\% on average, while adding 30 more templates reduces the $\chi^2$ by only $\sim$1\% on average. Hence, a larger number of templates was not considered due to its very limited effect. Since the reported $\sigma$ is based on the Set80 and the corresponding statistical uncertainty (see \S \ref{sec:bestfiterror}) is not very different from that of the individual template analysis (see \S \ref{sec:staterror}), we obtained the statistical uncertainties for each sample galaxy by applying the method described in \S\ref{sec:staterror} and \S\ref{sec:bestfiterror} for Set80 (see column 10 of the Table \ref{tab:results}). Assuming that the use of the Set80 eliminates the mismatch effect (see \S\ref{sec:mismatch}), one can take these values as the final uncertainty of the measured velocity dispersions. Being more conservative, in column 8 of Table \ref{tab:results}, we derived the final error of the measured $\sigma$ as the quadratic combination of both uncertainties (columns 10 \& 11). These values are used for further analysis in this paper and paper II (Kourkchi et al. 2011b).

We performed the same analysis to obtain the uncertainty of the measured radial velocities. The effect of the template mismatch on measured radial velocities is less than 1\%, because this measurement is only sensitive to the exact location of the absorption lines of the applied stellar templates which are slightly altered from one template to another. The final radial velocities and their corresponding uncertainties are obtained based on the Set80 run (columns 5 \& 6 of the Table \ref{tab:results}).

\begin{figure*}
\begin{center}
\includegraphics[width=10cm]{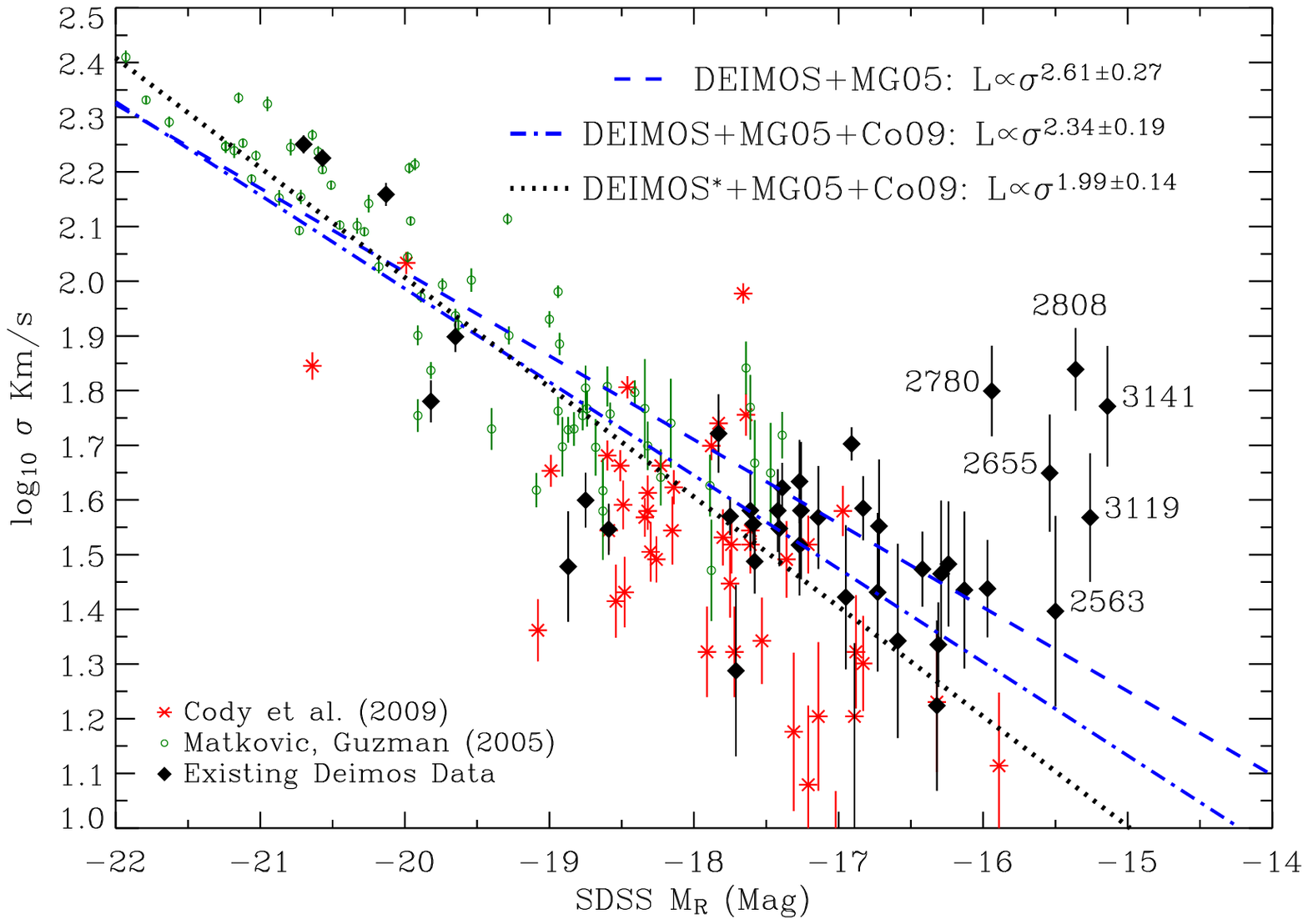}
\caption {The Faber-Jackson relation for three samples of dwarf galaxies in the Coma cluster. Filled black diamonds represent dwarf elliptical sample from DEIMOS/Keck observations. Red asterisks are dEs from Cody et al. (2009) and green open circles are data from Matkovi\'c \& Guzm\'an (2005). Except one point, the DEIMOS sample provides the only measurements at $M_R\geq -16$. The dashed line is the least-square orthogonal distance fit for DEIMOS+MG05 data while the dash-dotted line is the best orthogonal fit for all three samples. Of our sample, six of faintest galaxies are labelled with their GMP IDs. Dotted line shows the least-square orthogonal fir excluding the six galaxies fainter than $M_R=-16$ mag. See Table \ref{tab:faber_jackson} for detailed results.}
\label{fig:fjackson1}
\end{center}
\end{figure*}

\section{Results and Discussion}
\label{results}

We present the radial velocities, $V_r$, and velocity dispersions, $\sigma$, of 41 dwarf elliptical galaxies in the Coma cluster in Table~\ref{tab:results}. Of these galaxies, 12 have $S/N< 15$ per pixel in their spectra. The reported values are measured based on Set80, as explained in \S \ref{chap:velocities}.

We compare our measured $V_r$ and $\sigma$ values with MG05, MLKC02 and Co09 in Figures \ref{fig:cz} and \ref{fig:sig}. The mean difference between our $V_r$ values and those of MG05 is $\sim$8 $km~s^{-1}$. However when we include the $V_r$ measurements from Co09, the mean difference is $\sim$20 $km~s^{-1}$. Comparison of the velocity dispersion with MG05 (see Fig. \ref{fig:sig}) shows that for smaller values of velocity dispersion (i.e. $\sigma < 50 ~ km~s^{-1}$), our values are relatively smaller. Such discrepancy is likely due to the different methods of measurement, and/or to different templates used for measurement and/or to different wavelength regimes used for analysis. Since CaT lines are more prominent in dE galaxies than the absorption lines in the wavelength range $\sim$4200-5700 \AA, as used in studies of MG05 and Co09, our measurements of $V_r$ and $\sigma$ are more robust. Our analysis also includes measurements with 80 mixed stellar templates and detailed analysis of the uncertainties.

In addition, of 41 studied galaxies in this paper, 12 galaxies were observed twice using two different masks. In Table \ref{tab:masks:results}, for each of these galaxies we compare the measured velocity dispersions using both observed spectra. We also added both spectra of these galaxies to get better signal-to-noise ratios and then repeated the measurements. As seen, considering the statistical uncertainties, the measurements are consistent.

\begin{center}
\begin{table}
\renewcommand{\arraystretch}{0.9}\renewcommand{\tabcolsep}{0.12cm}
\caption{The measured velocity dispersion of the galaxies which were observed with two different masks. Columns 2 through 5 are the mask IDs also presented in Table \ref{tab:observations}. For these galaxies, we added both spectra of two masks to get the larger signal-to-noise ratios. In last column, the measured velocity dispersion of the added spectra is presented. The velocity dispersion are in terms of km s$^{-1}$ and the errors are based on the statistical uncertainties. As seen, the measured velocity dispersions for each galaxy are consistent within the error bars.}
\centering
\begin{tabular}{c|c c c c|c}
\hline

GMP &   Coma1-1 & Coma1-2 & Coma2-1 & Coma2-2  & Add \\

\hline
\hline

3424 &  &  & 27$\pm$7 & 26$\pm$7 & 27$\pm$5  \\
3223 & 16$\pm$9 & 29$\pm$8 &  &  & 22$\pm$5 \\
3209 & 32$\pm$3 & 29$\pm$5 &  &  & 30$\pm$2 \\
3119 & 40$\pm$7 & 32$\pm$7 &  &  & 37$\pm$5 \\
3080 &  &  & 8$\pm$9 & 13$\pm$9 & 9$\pm$5 \\
2983 & 23$\pm$6 & 28$\pm$5 &  &  & 27$\pm$3 \\
2877 &  &  & 32$\pm$8 & 30$\pm$7 & 29$\pm$5 \\
2808 &  &  & 75$\pm$11 & 71$\pm$9 & 69$\pm$8 \\
2755 & 22$\pm$6 & 31$\pm$7 &  &  & 27$\pm$4 \\
2718 &  &  & 33$\pm$5 & 29$\pm$6 & 30$\pm$4 \\
2605 & 41$\pm$9 & 32$\pm$8 &  &  & 36$\pm$6 \\
2571 &  &  & 17$\pm$4 & 14$\pm$4 & 17$\pm$3 \\
\hline
\end{tabular}
\label{tab:masks:results}
\end{table}
\end{center}

% \subsection{Faber Jackson $L-\sigma$ relation}
\subsection{Faber Jackson Relation}
\label{trends}

The relation between luminosity and velocity dispersion of bright elliptical galaxies was originally discovered by Faber \& Jackson (1976) and is expressed as $L\propto\sigma^\alpha$. For brighter ellipticals $\alpha$ is $\sim$4 while fainter galaxies exhibit a shallower slope. Davies et al. (1983) were the first to note that $\alpha$ changes from $\sim$4 to $\sim$2.4, for elliptical galaxies fainter than $M_R\simeq21.7$ mag. This result was also confirmed by Held et al. (1992) who found $\alpha=2.5$ for dE galaxies. These results were finally extended to lower luminosities by MG05 and Co09 who found $\alpha \simeq 2.0$. In this paper, we present data for 26 dE galaxies in the Coma cluster down to $M_R\simeq-15$ mag. Using the orthogonal distance regression, we derive the F-J relation to be $L\propto\sigma^{2.34\pm0.19}$ for entire sample, and $L\propto\sigma^{1.99\pm0.14}$ for galaxies brighter than $M_R\simeq-16$ mag.

\begin{figure*}
\begin{center}
\hspace{-0.5cm}
\includegraphics[width=16.5cm]{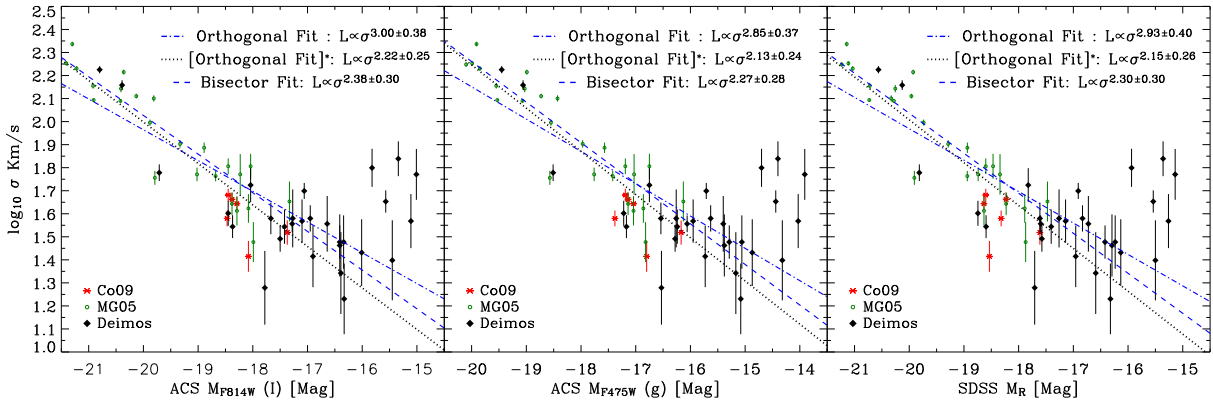}
\caption {Same as the Figure \ref{fig:fjackson1}. The Faber-Jackson relation for three samples of dwarf galaxies which lie in the footprint of the HST/ACS Coma Treasury survey (Carter et al. 2008). Left panel shows the FJ relation for this sample using magnitudes based on SDSS data. For left and middle panels, magnitudes are derived from the Treasury HST/ACS images in the F814W (I-band) and F475W (g-band) filters (Paper II). Dashed line is the least-square bisector fit and dash-dotted line is obtained using the orthogonal distance regression. Dotted line shows the orthogonal fit for galaxies brighter than $M_R=-16$ mag. We see no colour dependence of the Faber-Jackson within the error bars.}
\label{fig:fjackson2}
\end{center}
\end{figure*}

We show the $L-\sigma$ relation for the galaxies in our sample in Figures \ref{fig:fjackson1} and \ref{fig:fjackson2}, where we also include the sample of Co09 and MG05. We excluded the spiral galaxies from Co09 when deriving the F-J relation. $L-\sigma$ correlation coefficient is -83\%, at a significance of 99.99\% (i.e., only a 0.01\% chance that the correlation is due to the random scatter). We perform linear fits via two methods, the least square bisector fit and the orthogonal distance regression. Both methods incorporate the errors in our measurements and those reported in the literature.
In orthogonal distance regression, the orthogonal distance of data points from the line is minimized. The magnitude of all galaxies in Figure \ref{fig:fjackson1} and the left panel of Figure \ref{fig:fjackson2} are derived from SDSS DR7. To transform SDSS $ugriz$ magnitudes into Johnson-Cousins R magnitude, we used the mean value of the following relations presented by Lupton 2005\footnote{For various transformations between SDSS $ugriz$ magnitudes to Johnson-Cousins $UBVRI$ system see Jordi et al. (2006), Ivezi\'c et al. (2007) and \\{\tt \it http://www.sdss.org/dr6/algorithms/sdssUBVRITransform.html}}:

%\begin{center}
\begin{table*}
%\centering
\begin{minipage}{135mm}
\renewcommand{\arraystretch}{0.9}\renewcommand{\tabcolsep}{0.12cm}
\caption{Least square fit to find $L-\sigma$ relation. (1) The sample galaxies used in regression. (2) Surveys-Filters for magnitude of the sample galaxies. (3) The order of regression. For example, in the first row residuals in $log\sigma$ are minimized. (4,5) The slope and intercept of $M=A~log\sigma+B$ and their 1$\sigma$ uncertainties. (6) $\alpha$ in the Faber-Jackson relation ($L\propto\sigma^\alpha$). We have used the data with $S/N>15$ per \AA of MG05 and all dEs of Co09.}
\label{tab:faber_jackson}
\begin{tabular}{ c c c c c c }
\hline
Sample Galaxies &  Filter & Regression  &   Slope (A) &   Intercept (B)  &   $\alpha$  \\  
(1)  & (2) & (3) & (4) & (5)  & (6)  \\ \hline \hline

DEIMOS+MG05 &  SDSS-R &  $log\sigma$  & $-6.59\pm0.67$  &  $-6.72\pm2.17$  &  $2.64\pm0.27$ \\
DEIMOS+MG05 &  SDSS-R &  $M_R$  & $-4.55\pm0.38$  &  $-10.45\pm0.76$  &  $1.81\pm0.15$ \\
DEIMOS+MG05 &  SDSS-R &  $Bisector$  & $-5.57\pm0.55$  &  $-8.59\pm1.63$  &  $2.23\pm0.22$ \\
DEIMOS+MG05 &  SDSS-R &  $Orthogonal$  & $-6.52\pm0.67$  &  $-6.84\pm2.19$  &  $2.61\pm0.27$ \\
DEIMOS$^*$+MG05 & SDSS-R &  $Orthogonal$  & $-5.41\pm0.49$  &  $-9.02\pm2.02$  &  $2.16\pm0.20$ \\
\\
DEIMOS+MG05+Co09 & SDSS-R &  $Bisector$  & $-4.85\pm0.40$  &  $-10.12\pm1.31$  &  $1.94\pm0.16$ \\
DEIMOS+MG05+Co09 & SDSS-R &  $Orthogonal$  & $-5.85\pm0.49$  &  $-8.37\pm1.79$  &  $2.34\pm0.19$ \\
DEIMOS$^*$+MG05+Co09 & SDSS-R &  $Orthogonal$  & $-4.98\pm0.35$  &  $-10.00\pm1.58$  &  $1.99\pm0.14$ \\
\\
DEIMOS+MG05+Co09 & CFHTLS-i &  $Bisector$  & $-4.73\pm0.38$  &  $-10.33\pm1.27$  &  $1.89\pm0.15$ \\
DEIMOS+MG05+Co09 & CFHTLS-i &  $Orthogonal$  & $-5.61\pm0.45$  &  $-8.81\pm1.73$  &  $2.25\pm0.18$ \\
DEIMOS$^*$+MG05+Co09 & CFHTLS-i &  $Orthogonal$  & $-4.82\pm0.33$  &  $-10.29\pm1.54$  &  $1.93\pm0.13$ \\
\\
\hline
$[$DEIMOS$^*$+MG05+Co09$]$ & HST/ACS-F814W (I) &  $Orthogonal$  & $-5.55\pm0.63$  &  $-8.90\pm2.49$  &  $2.22\pm0.25$ \\
\\
$[$DEIMOS$^*$+MG05+Co09$]$ & HST/ACS-F475W (g) &  $Orthogonal$  & $-5.33\pm0.61$  &  $-8.02\pm2.35$  &  $2.13\pm0.24$ \\
\\
$[$DEIMOS$^*$+MG05+Co09$]$ & SDSS-R &  $Orthogonal$  & $-5.37\pm0.64$  &  $-9.21\pm2.64$  &  $2.15\pm0.26$ \\

\hline
\end{tabular}
All galaxies in $[$DEIMOS+MG05+Co09$]$ sample have HST/ACS data. DEIMOS$^*$ represents the DEIMOS sample excluding galaxies fainter than $M_R=-16$ mag. 
\end{minipage}
\end{table*}
%\end{center}

\begin{equation}
R = r - 0.1837*(g - r) - 0.0971 ,
\end{equation}

\begin{equation}
R = r - 0.2936*(r - i) - 0.1439 .
\end{equation}

The results of the least square fits are available in Table \ref{tab:faber_jackson}. We run the same analysis for the same sample using the CFHT Megacam i-band data. We find that our FJ relation slope is consistent with those of Co09 and MG05 within the errors, for both linear fitting methods. The least square bisector method yields $L_R\propto\sigma^{1.94\pm0.16}$, while we derive $L_R\propto\sigma^{2.34\pm0.19}$ via the orthogonal regression method. Using the CFHT i-band data, the best fits of FJ relation are consistent with those derived from SDSS data (see Table \ref{tab:faber_jackson}). The behaviour at the faint end of the FJ relation is uncertain as many of our galaxies in this magnitude range ($M_R>-16$) have $S/N<15$ per pixel. However our results imply that the slope of the FJ relation may change towards higher velocity dispersions at this faint end. Excluding galaxies fainter than $M_R\simeq-16$, FJ is $L_R\propto\sigma^{1.99\pm0.14}$.

Next, we use photometry from the HST/ACS Coma Treasury survey (Carter et al. 2008) to derive the FJ relation. Out of 41 galaxies in our sample, 28 have available HST/ACS magnitudes, together with 43 galaxies from MG05 and Co09 samples. In Figure \ref{fig:fjackson2} and Table \ref{tab:faber_jackson}, we show the best linear fits of the FJ relation using HST/ACS F814W (I-band) and F475W (g-band). We find that the FJ relation shows no discrepancy on colour.

Following all above analysis, we noted that for one of our sample galaxies (GMP 3308), the SDSS magnitudes are almost one magnitude fainter than what is derived from CFHT and HST images. Therefore, we modified the SDSS R magnitude of this galaxy by constructing a linear trend between the CFHT i-band and SDSS R magnitudes of all sample galaxies.

\begin{figure*}
\begin{center}
\hspace{0.9in}
\includegraphics[width=17cm]{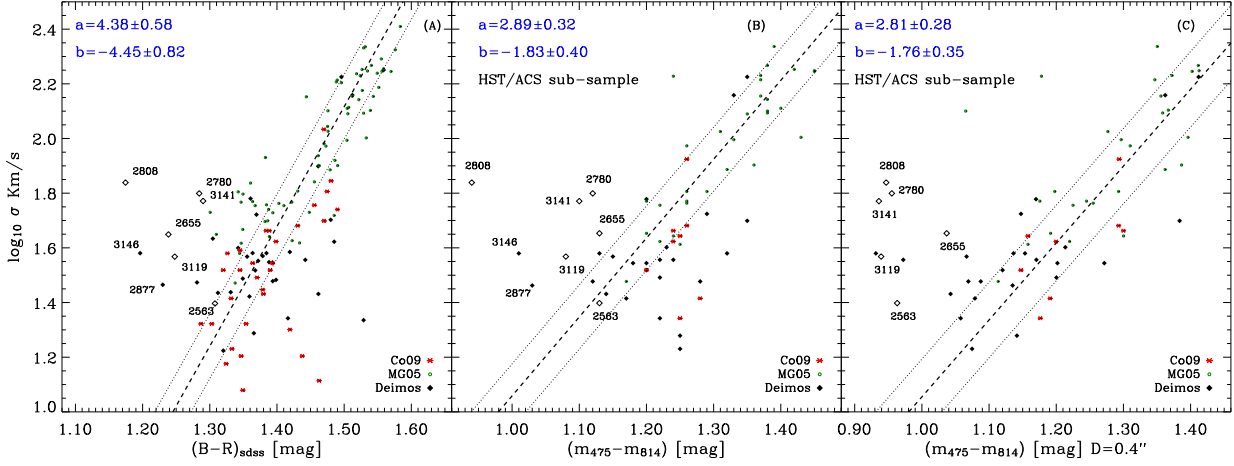}
\caption {Colour-$\sigma$ relation for three samples of dwarf ellipticals in the Coma cluster. Filled black diamonds represent dEs from our DEIMOS/Keck sample. Open diamonds represent faintest dEs of our sample which are labelled with their GMP IDs. Red asterisks are dEs from Co09 and green open circles are from MG05. Dashed line shows the best linear orthogonal fit for brighter galaxies, $M_R>-17$ mag and dash-dotted lines represent 1$\sigma$ uncertainty along the vertical axis. In all panels, `a' and `b' are slope and intercept of the fitted lines, respectively. Panel (A) shows the C-$\sigma$ relation using the magnitudes from SDSS DR7. Panel (B) shows the C-$\sigma$ relation for a sub-sample of dEs for which their HST/ACS magnitudes in F475W (g) and F814W (I) bands are available. Panel (C) is the same as panel (B) except for the colour of galaxies which are calculated within the aperture with diameter of $0.4''$ (i.e. ~185 pc at the location of the Coma cluster).}
\label{fig:csigma}
\end{center}
\end{figure*}

\subsection{Colour-$\sigma$ relation}
\label{col-sigma}

The relation between dynamical mass of galaxies and their stellar population can be studied using the colour-$\sigma$ diagram. MG05 showed a well-defined colour-$\sigma$ relation for faint early type galaxies brighter than $M_R=-17.5$ mag. In panel (A) of Figure \ref{fig:csigma}, we extend the relation to fainter dEs ($M_R<-15$ mag) which are located at the blue end of the diagram and are more scattered around the the fitted line for brighter galaxies (i.e. for $M_R>-17$ mag the relation is $B-R=(0.23\pm0.03)log\sigma+(1.02\pm0.23)$). The scatter of dEs around the colour-$\sigma$ relation could be due to their different formation history, age and metallicity. To calculate the colour of the galaxies, B-R, we used the SDSS data and Lupton (2005) transformation functions to derive the Johnson B and R magnitudes. Panel (B) of Figure \ref{fig:csigma} shows the same relation using $m_{475}-m_{814}$ colour of the galaxies in the HST/ACS field.

The dependence of the internal velocity dispersion of dEs and their central activity is investigated in panel (C) of Figure \ref{fig:csigma}. Central colour of the galaxies, for which we have the HST/ACS images, are calculated within the aperture with diameter of $0.4''$ (i.e. ~185 pc at the location of the Coma cluster). We found a linear trend between the velocity dispersion of dEs and their central colours, except for very faint galaxies which have bluer core, $(m_{475}-m_{814})<1.0$.

\begin{center}
\begin{table}
\caption{The results of $\sigma$ measurements for 8 blue galaxies based on different template libraries.}
\centering
\begin{tabular}{ c | c c c  }
\hline
GMP &  \multicolumn{3}{c}{$\sigma$ ($km~s^{-1}$)}    \\  
ID  &  Set80  &  FGK-lib & F-lib   \\ 
\hline \hline
3146  &  38  &  37  &  35  \\
3141  &  59  &  60  &  61  \\
3119  &  37  &  35  &  26  \\
2877  &  29  &  27  &  26  \\
2808  &  69  &  39  &  38  \\
2780  &  63  &  63  &  31  \\
2655  &  45  &  45  &  42  \\
2563  &  25  &  25  &  28  \\
\hline
\end{tabular}
\label{tab:sig-lib}
\end{table}
\end{center}

\subsection{Remarks on faint galaxies}
\label{remarks}

Of six faintest galaxies from Deimos sample, GMP2808, GMP2655, GMP3119, GMP2780 and GMP3141 are bluer than all other sample galaxies (see Figure \ref{fig:csigma}) and have larger velocity dispersion with respect to the L-$\sigma$ relation of brighter galaxies (see Figure \ref{tab:faber_jackson}).

In order to test the effect of stellar population on measured velocity dispersions, we have repeated the $\sigma$ measurement using two sets of stellar templates. First set consists of 10 F-stars with different spectral types and metallicities (hereafter F-lib). Second set includes both F-lib and the G/K-stars of Table \ref{tab:ftemplates} (hereafter FGK-lib). The results of $\sigma$ measurements based on Set80, FGK-lib and F-lib are compared in Table \ref{tab:sig-lib}. The analysis based on FGK-lib shows no changes in results except for GMP2808 which is the bluest galaxy in our sample. The measured $\sigma$ of GMP3119 and GMP2808 in F-lib run is smaller than those of Set80 and FGK-lib runs, however the resulting $\chi^2$ is higher for F-lib run.
The maximum change in measured $\sigma$ for all other galaxies in our sample in FGK-lib run is less than 2 $km~s^{-1}$.  As a results, the measured $\sigma$ values in Table \ref{tab:results} are independent of the stellar population except for the bluest galaxy, for which blue stellar templates should be used for $\sigma$ measurement.

Recent gas-rich mergers in blue dEs may be responsible for their deviation from the FJ relation. Supernova driven winds can expel the luminous matter of these galaxies and shift them to the faint end of the FJ diagram. Moreover, the resulting starburst due to the inward gas transportation during the merger, produces the central excess light compared to the extrapolated S\'{e}rsic function of the galaxy outer region (Kormendy et al. 2009). The correlation between the central extra light and deviation of faint dEs from the scaling relations of brighter ellipticals is studied in paper II.

\section{Summary}
\label{conclusions}

We have observed a sample of $\sim$50 dwarf elliptical galaxies in the core of Coma cluster of which for 41 dwarfs we measured their internal velocity dispersion. For 26 galaxies, we presented the velocity dispersion for the first time. Our current study extends the relation one magnitude fainter than previous studies in Coma. We performed a comprehensive analysis to find the source of the uncertainties when measuring the velocity dispersion. We found that the main uncertainty may arise due to the mismatch of the stellar templates. This kind of uncertainty is less than 20\% of the measured velocity dispersion for $\sigma\sim50~km~s^{-1}$. The template mismatch uncertainty decreases for higher velocity dispersions and is reduced to $\sim5\%$ for $\sigma\gea100~km~s^{-1}$. We find greater velocity dispersions when using templates with lower stellar metallicities. The effect of template mismatch is reduced with the use of multiple stellar templates and the optimization of the template weights to reproduce each sample galaxy spectrum. In addition, using the multiple stellar templates removes the dependency of the measurements on galaxy and/or templates metallicity. Casting our data points and those from the literature, we get L-$\sigma$ relation as $L\propto\sigma^{1.99\pm0.14}$. We noticed that fainter dwarfs show a departure from the FJ relation of the brighter ellipticals which indicates that they have higher velocity dispersion than what is predicted by L-$\sigma$ linear trend. In paper II, we will present the combined study of the photometric and spectroscopic properties of this sample, focusing on the fundamental and the photometric plane of the galaxies.
%==============================================================
\section*{Acknowledgments}

The authors wish to recognize and acknowledge the very significant cultural role and reverence that the summit of Mauna Kea has always had within the indigenous Hawaiian community.  We are most fortunate to have the opportunity to conduct observations from this mountain. DC and AMK acknowledge support from the Science and Technology Facilities Council, under grant PP/E/001149/1. EMQ acknowledges support from the SMES for a FPI PhD fellowship through the research project AYA2007-67752-C03-01. EK would like to thank Paul Westoby and Michele Cappelari for their helpful comments to run pPXF. EK also acknowledges financial supports from Sharif University of Technology.
This research also used the facilities of the Canadian Astronomy Data Centre operated by the National Research Council of Canada with the support of the Canadian Space Agency. 
%===========================================

%===========================================

\bsp

\label{lastpage}


\begin{thebibliography}{}


\bibitem[Aaronson (1983)]{Aa03} Aaronson, M.\ 1983, ApJL, 266, L11
\bibitem[Adami et al. (2006)]{Ad06} Adami, C., et al.\ 2006, A\&A, 451, 1159
\bibitem[Bender et al. (1992)]{Ben02} Bender, R., Burstein, D. \& Faber, S.M.\ 1992, ApJ, 399, 462
\bibitem[Cappellari \& Emsellem (2004)]{CE04} Cappellari, M. \& Emsellem, E.\ 2004, PASP, 116, 138
\bibitem[Carrera et al. (2007)]{Ca07} Carrera, R., Gallart, C., Pancino, E. \& Zinn, R. \ 2007, AJ, 134, 298
\bibitem[Carter et al. (2008)]{Ca08} Carter, D., et al.\ 2008, ApJS, 176, 424 
\bibitem[Cenarro et al. (2003)]{Ce03} Cenarro, A. J., et al.\ 2003, MNRAS, 339, 12
\bibitem[Cody et al. (2009)]{Co09} Cody. A.M., Carter, D., Bridges, T. J., Mobasher, B. \& Poggianti, B. M.\ 2009, MNRAS, 396, 1647 (Co09)
\bibitem[Davis et al. (2003)]{Da03} Davis, M., Faber, S.M., Newman, J.A., et al.\ 2003, Proc SPIE, 4834, 161
\bibitem[De Rijcke et al. (2006)]{Dr06} De Rijcke, S., Prugniel, P., Simien F., \& Dejonghe, H., \ 2006. MNRAS, 369, 1321
\bibitem[Dekel \& Silk (1986)]{DS86} Dekel. A., \& Silk, J.\ 1986, ApJ, 303, 39
\bibitem[D\'iaz et al. (1989)]{DTT89} D\'iaz, A I., Terlevich, E., Terlevich, R.\ 1989, MNRAS, 239, 325
\bibitem[Djorgovski \& Davis (1987)]{DD87} Djorgovski, S. \& Davis, M.\ 1987, ApJ, 313, 59
\bibitem[Dressler et al. (1987)]{Dr87} Dressler, A., Lynden-Bell, D., Burstein, D., Davies, R.L., Faber, S.M., Terlevich, R. \& Wegner, G.\ 1987, ApJ, 313, 42
\bibitem[Faber et al. (2003)]{Fab03} Faber, S.M. et al.\ 2003, Proc SPIE, 4841, 1657 
\bibitem[Faber \& Jackson (1976)]{FJ76} Faber, S. M., \& Jackson, R. E., 1976, ApJ, 204, 668
\bibitem[Foster et al. (2010)]{Fo10} Foster, C., Forbes, D. A., Proctor, R. N., Strader, J., Brodie, J. P. \& Spitler,  L. R. \ 2010, AJ, 139, 1566
\bibitem[Geha et al. (2002)]{Ge02} Geha, M., Guhathakurta, P. \& van der Marel, R. P. \ 2002., AJ, 124, 3073
\bibitem[Godwin \& Metcalfe \& Peach (1983)]{GMP} Godwin J., G., Metcalfe, N., \& Peach, J., V. \ 1983, MNRAS, 202, 113
\bibitem[Graham \& Guzman (2003)]{GG03} Graham, A.W. \& Guzm\'an, R. \ 2003, AJ, 125, 2936 (GG05)
\bibitem[Held et al. (1992)]{He92} Held, E. V., de Zeeuw, T., Mould, J., \& Picard, A., 1992, AJ, 103,851
\bibitem[Ivezic et al. (2007)]{Iv07} Ivezi\'c, {\v Z}., et al. \ 2007, ASPC, 364, 165
\bibitem[Jordi et al. (2006)]{Jo06} Jordi, K., Grebel, E. K. \& Ammon, K. \ 2006, A\&A, 460, 339
\bibitem[Jorgensen et al. (1996)]{Jor96} J{\o}rgensen, I., Franx, M. \& Kjaergaard, P.\ 1996, MNRAS, 280, 167
\bibitem[Jorgensen et al. (1992)]{Jor92} J{\o}rgensen, U. G., Carlsson, M. \& Johnson, H. R.\ 1992, A\&A, 254, 258
\bibitem[Kormendy et al. (2009)] {Ko09} Kormendy, J., Fisher, D.B., Cornell, M.E. \& Bender, R.\ 2009, ApJS, 182, 216
\bibitem[Kourkchi et al. (2011b)]{Ko11b} Kourkchi, E., Khosroshahi, H. G., Carter, D. \& Mobasher, B.\ 2011b (submitted to MNRAS; Paper II)
\bibitem[Lejeune et al. (1997)]{LCB97} Lejeune, Th., Cuisinier, F., Buser, R. \ 1997, A\&AS, 125, 229
\bibitem[Mateo (1998)]{Mat98} Mateo, M.\ 1998, ARA\&A, 36, 435
\bibitem[Mallik (1994)]{ML94} Mallik, S. V. \ 1994, A\&AS, 103, 279	
\bibitem[Marzke et al. (2011)]{Mr11} Marzke et al.\ 2011 (in preparation)	
\bibitem[Matkovi\'c \& Guzm\'an (2005)]{MG05} Matkovi\'c, A. \& Guzm\'an, R.\ 2005, MNRAS, 362, 289
\bibitem[Michielsen et al. (2007)] {MKP07} Michielsen, D. et al.\ 2007, ApJ, 670, 101
\bibitem[Moore et al. (2002)] {MLKC02} Moore, S.A.W., Lucey, J.R., Kuntschner, H. \& Colless, M.\ 2002, MNRAS, 336, 382 (MLKC02)
\bibitem[Pritchet (1978)] {PR78} Pritchet, C. \ 1978 ApJ, 221, 507
\bibitem[Schulz et al. (2002)]{SAMF02} Schulz, J., Fritze-v. Alvensleben, U., M\"{o}ller, C. S. \& Fricke, K. J. \ 2002, A\&A, 392, 1-11
\bibitem[Smith \& Drake (1990)] {SD90} Smith, G., Drake, J. J. \ 1990, A\&A, 231, 125
\bibitem[Starkenburg et al. (2010)]{St10} Starkenburg, E., et al. \ 2010, A\&A, 513A, 34
\bibitem[Toloba et al. (2011)]{TOL11} Toloba, E., Boselli, A., Cenarro, A. J., Peletier, R. F., Gorgas, J., Gil de Paz, A. \& Mu\~noz-Mateos, J. C., \ 2011, A\&A, 526, 114
\bibitem[Tonry \& Davis (1979)]{TD79} Tonry, J. \& Davis, M. \ 1979, AJ, 84, 1511
\bibitem[Valdes et al. (2004)]{Val04} Valdes, F., Gupta, R., Rose, J.A., Singh, H.P. \& Bell, D.J. \ 2004, ApJS, 152, 251
\bibitem[Wilkinson et al. (2002)]{Wi02} Wilkinson, M. I., Kleyna, J., Evans, N. W. \& Gilmore, G. \ 2002, MNRAS, 330, 778
\bibitem[Yoshii \& Arimoto (1987)]{YA87} Yoshii, Y. \& Arimoto, N. \ 1987, A\&A, 188, 13
\bibitem[Zhou, Xu (1991)]{ZX91} Zhou, Xu\ 1991, A\&A, 248, 367


\end{thebibliography}
\end{document}